\documentclass[twocolumn]{aastex62}

\usepackage{amsmath}
\usepackage[varg]{txfonts}
\usepackage{natbib}
\usepackage{xspace}
\usepackage{graphicx}

\hypersetup{linkcolor=red,citecolor=cyan,filecolor=green,urlcolor=magenta}

\newcommand{\nnhp}{\mbox{N$_2$H$^+$}\xspace}

\newcommand{\nhhh}{\mbox{NH$_3$}\xspace}

\renewcommand{\micron}{\mbox{$\mathrm{\mu m}$}\xspace}
\newcommand{\um}{\mbox{\micron}\xspace}

\newcommand{\Tkin}{\mbox{$T_{\mathrm{kin}}$}\xspace}

\newcommand{\sig}{\mbox{$\sigma$}\xspace}
\newcommand{\signt}{\mbox{$\sigma_{\rm nt}$}\xspace}
\newcommand{\sigobs}{\mbox{$\sigma_{\rm obs}$}\xspace}
\newcommand{\sigchan}{\mbox{$\sigma_{\rm chan}$}\xspace}
\newcommand{\kmspc}{\mbox{$\mathrm{km~s^{-1}~pc^{-1}}$}\xspace}
\newcommand{\kms}{\mbox{$\mathrm{km~s^{-1}}$}\xspace}

\newcommand{\Vlsr}{\mbox{$V_{\mathrm{lsr}}$}\xspace}
\newcommand{\Vobs}{\mbox{$V_{\mathrm{obs}}$}\xspace}
\newcommand{\Vchan}{\mbox{$V_{\mathrm{chan}}$}\xspace}
\newcommand{\Ms}{\mbox{$M_{\sun}$}\xspace}
\newcommand{\Msol}{\Ms}

\newcommand{\mgas}{\mbox{$m_{\rm gas}$}\xspace}
\newcommand{\mnhhh}{\mbox{$m_{\rm NH_3}$}\xspace}
\newcommand{\cs}{\mbox{$c_{\rm s}$}\xspace}
\newcommand{\kb}{\mbox{$k_{\rm B}$}\xspace}
\newcommand{\sigsigobs}{\mbox{$\sigma_{\sigma_{\rm obs}}$}\xspace}
\newcommand{\sigTkin}{\mbox{$\sigma_{T_{\rm kin}}$}\xspace}
\newcommand{\sigMach}{\mbox{$\sigma_{\mathcal{M}}$}\xspace}
\newcommand{\velgradmag}{\mbox{$\| \nabla \Vlsr \|$}\xspace}

\turnoffedit

\received{2018 September 30}
\accepted{2018 December 18}
\submitjournal{ApJ}

\shorttitle{Ammonia kinematics in IRDC G035.39}
\shortauthors{Sokolov et al.}

\begin{document}

\title{\large \textbf{Multicomponent kinematics in a massive filamentary IRDC}}

\correspondingauthor{Vlas Sokolov}
\email{vlas.sokolov@mpg-alumni.de}

\correspondingauthor{Jaime E. Pineda}
\email{jpineda@mpe.mpg.de}

\author[0000-0002-5327-4289]{Vlas Sokolov}
\affil{Max Planck Institute for Extraterrestrial Physics, Gie{\ss}enbachstra{\ss}se 1, D-85748 Garching bei M{\"u}nchen, Germany}

\author[0000-0002-7237-3856]{Ke Wang}
\affiliation{European Southern Observatory (ESO) Headquarters, Karl-Schwarzschild-Str. 2, 85748 Garching bei M\"{u}nchen, Germany}
\affiliation{Kavli Institute for Astronomy and Astrophysics, Peking University, 5 Yiheyuan Road, Haidian District, Beijing 100871, China}

\author[0000-0002-3972-1978]{Jaime E. Pineda}
\affiliation{Max Planck Institute for Extraterrestrial Physics, Gie{\ss}enbachstra{\ss}se 1, D-85748 Garching bei M{\"u}nchen, Germany}

\author[0000-0003-1481-7911]{Paola Caselli}
\affiliation{Max Planck Institute for Extraterrestrial Physics, Gie{\ss}enbachstra{\ss}se 1, D-85748 Garching bei M{\"u}nchen, Germany}

\author{Jonathan D. Henshaw}
\affiliation{Max Planck Institute for Astronomy, K\"{o}nigstuhl 17, D-69117 Heidelberg, Germany}

\author[0000-0003-0410-4504]{Ashley T. Barnes}
\affiliation{Argelander Institute for Astronomy, University of Bonn, Auf dem H\"{u}gel 71, 53121 Bonn, Germany}

\author[0000-0002-3389-9142]{Jonathan C. Tan}
\affiliation{Department of Astronomy, University of Florida, Gainesville, FL, 32611, USA}
\affiliation{Department of Physics, University of Florida, Gainesville, FL, 32611, USA}

\author[0000-0003-0348-3418]{Francesco Fontani}
\affiliation{INAF-Osservatorio Astrofisico di Arcetri, Largo E. Fermi 5, I-50125 Firenze, Italy}

\author[0000-0003-4493-8714]{Izaskun Jim\'enez-Serra}
\affiliation{Centro de Astrobiolog\'ia (CSIC/INTA), Ctra de Torrej\'on a Ajalvir, km 4, 28850 Torrej\'on de Ardoz, Spain}

\begin{abstract}
\edit1{To probe the initial conditions for high-mass star and cluster formation,} we investigate the properties of dense filaments within the infrared dark cloud G035.39--00.33 (IRDC G035.39)
\edit1{in a combined Very Large Array (VLA) and the Green Bank Telescope (GBT) mosaic tracing the \nhhh (1,1) and (2,2) emission down to 0.08 pc scales.}
Using agglomerative hierarchical clustering \edit1{on multiple line-of-sight velocity component fitting results}, we identify seven extended velocity-coherent components in our data, likely representing spatially coherent physical structures\edit1{, some exhibiting complex gas motions.}
The velocity gradient magnitude distribution peaks at its mode of 0.35 \kmspc and has a long tail extending into higher values of $1.5 - 2$ \kmspc, and is generally consistent with those found toward the same cloud in other molecular tracers and with
the values found towards nearby low-mass dense cloud cores at the same scales.
Contrary to observational and theoretical expectations, we find the \edit1{non-thermal ammonia} line widths to be systematically narrower (by about 20\%) than those of \nnhp ($1-0$) line transition observed with similar resolution.
If the observed ordered velocity gradients represent the core envelope solid-body rotation, we estimate the specific angular momentum to be about $2 \times 10^{21}~\mathrm{cm^{2}~s^{-1}}$, similar to the low-mass star-forming cores.
Together with the previous finding of subsonic motions in G035.39, our results demonstrate high levels of similarity between kinematics of a high-mass star-forming IRDC and the low-mass star formation regime.
\end{abstract}

\keywords{ISM: kinematics and dynamics --- ISM: clouds --- stars: formation --- ISM: individual objects: G035.39-00.33}

\section{Introduction}\label{sec:intro}

The formation of stars from the diffuse ISM is a multi-stage process of gas channeling and mass assembly. In the well-established low mass star formation picture, individual stars are formed from dense cores, which become gravitationally unstable once enough mass has been accumulated from their surroundings. While this process is thought to be relatively well understood, the formation of high-mass stars and star clusters occurs more rapidly and is a more dynamic process, requiring efficient gas channeling onto the protostellar sources, and its initial conditions are still not very well understood~\citep[e.g.,][]{tan+2014}.

Observations of the earliest stages of massive star and cluster formation are greatly hindered by the high extinction of, and large distances to, high-mass star-forming regions. Thus, resolving the physical scales of the massive starless cores typically requires interferometric observations in radio or submillimeter wavelengths.
The initial conditions of this process, such as density and temperature profiles on the core scales (${\sim} 0.1$ pc), turbulence levels, accretion from the surrounding envelopes, magnetic field strengths, or the angular momentum of the core, are inherited from the more extended scales of clouds or protostellar clumps (${\sim} 1$ pc), and are often poorly constrained by observations.
As these conditions directly regulate the subsequent fragmentation and infall of the star-forming material~\citep{tan2017, motte+2018}, high-angular resolution observations of the dense IR-dark gas with little evidence of disruptive feedback can provide improved constraints on the theoretical models and numerical simulations aiming to recreate the formation of high-mass stars and star clusters. 

Large-scale infrared surveys of the Galactic plane~\citep{perault+1996, egan+1998} have revealed Infrared Dark Clouds, or IRDCs, which are the most promising candidates to host the earliest stages of massive star and cluster formation~\citep{rathborne+2006}. Often appearing as dense extinction ridges, tracing higher-density spines of their parent GMCs~\citep{hernandez+tan2015, schneider+2015}, IRDCs appear as cold ($10-15$ K), filamentary, high-column-density ($10^{23}-10^{25}$ cm$^{-2}$) and high-mass~\citep[$10^{3-5}$ \Msol,][]{kainulainen+tan2013} extinction features, typically hosting a number of high-mass cores~\citep{rathborne+2006}. IRDCs vary greatly in their masses, densities, and star-forming activity~\citep{kauffmann+pillai2010}.
\cite{rathborne+2006} have selected 38 darkest clouds representative of the earliest stages of massive star formation from the~\cite{simon+2006} sample, which contains over 10,000 IRDCs, and shown that the clouds are further fragmented into high-mass cores with a median mass of 120 \Msol. \cite{butler+tan2009} studied mid-infrared extinction of 10 IRDCs from the~\cite{rathborne+2006} sample, selecting the clouds that are nearby, massive, 
and show relatively simple diffuse emission. \cite{butler+tan2012} identified a number of compact ($R \sim 0.1-0.2$ pc) dense cores embedded in these ten IRDCs and discussed their star-forming potential.

In this study, we investigate the physical properties of dense gas connecting the cloud to core scales (from the IRDC size of ${\sim} 6$ pc down to \edit1{0.08} pc) for one of ten clouds in~\cite{butler+tan2009, butler+tan2012}, IRDC G035.39--00.33.
It is seen as a highly filamentary feature at mid-infrared wavelengths \citep[e.g.,][]{butler+tan2009} and belongs to a larger GMC complex \citep{hernandez+tan2015}.
By combining available NIR and MIR extinction measurements, \cite{kainulainen+tan2013} have estimated the mass of G035.39 to be 16700 \Msol.
The cloud temperature is relatively cold \citep[${\sim} 15$ K from dust emission measurements,][]{nguyen_luong+2011}, with high CO depletion and large deuteration fractions from single dish observations indicating that the cloud is governed by cold chemistry with little evidence of stellar feedback \citep{hernandez+2011, jimenez-serra+2014, barnes+2016}.
\cite{jimenez-serra+2010} have discovered extended SiO (2-1) emission throughout the northern part of the cloud, suggesting that its origin may arise from a slow shock from the formation of the IRDC \citep[a possibility later explored by][]{henshaw+2013, liu+2018} or from the stellar feedback driven by embedded YSOs \citep[suggested to be only partially responsible for the SiO emission][]{nguyen_luong+2011}.
Despite its starless appearence on large, pc-scales, \cite{nguyen_luong+2011} find a population of protostellar cores within G035.39, nine of which are massive enough ($> 20$ \Msol) to form intermediate to high-mass stars.
On smaller scales, \cite{henshaw+2014} have studied the kinematics of the dense gas in the northern part of the IRDC, resolving narrow dense filaments in it with a velocity pattern suggesting a dynamical interaction between the filaments and the dense cores in the cloud.
Previously, \cite{sokolov+2017} have probed the temperature structure of the IRDC Green Bank Telescope (GBT) observations and found evidence for a collective gas heating from the embedded protostars in G035.39, and \cite{sokolov+2018} presented the Very Large Array (VLA) findings of narrow line widths in the IRDC G035.39 and discussed its implications for the high-mass star-forming regime.

In this paper, we present a detailed analysis of the combined Very Large Array (VLA) and Green Bank Telescope (GBT) ammonia observations of the whole G035.39 cloud. The paper is structured as follows. In \S\ref{sec:data-reduction} we summarize the data reduction, our strategy for VLA and GBT dataset combination, and present an overview of its outcome. In \S\ref{sec:results} we describe the separation of the data into individual components, and present the spectral line fitting results for the largest among them. In \S\ref{sec:discussion} a quantitative comparison with similar studies of G035.39 is made, the implications of cloud kinematics for velocity flows around the protostellar and continuum sources are made, and the stability of the individual components is being discussed. Finally, a summary of our main results comprise section \S\ref{sec:conclusions}.

\section{Data Reduction}\label{sec:data-reduction}

The Karl G. Jansky Very Large Array (VLA) observations (project AW776; PI: Ke Wang) in compact D-configuration were carried out on May 8, 2010, mapping the \nhhh (1,1) and (2,2) inversion lines in the K-band across the whole G035.39 cloud in a five-point mosaic.
The two ammonia lines were mapped in two consecutive sessions, to cover the full hyperfine line structure of both transitions with the first generation VLA correlator \edit1{at a channel separation} of 15.625 kHz. \edit1{The effective spectral resolution is 18.75 kHz, which is the FWHM of the sinc spectral response function of the correlator.}
The data were calibrated on the quasars J1851+005 (gain), J2253+1608 (bandpass), and 3C48 (flux) within the CASA data reduction package. The data presented have been previously described in~\cite{sokolov+2018}.

We have deconvolved the calibrated visibilities in CASA, using the \textsc{tclean} task, with the multi-scale CLEAN algorithm~\citep{cornwell2008}, following the Briggs weighting with robust parameter set to 0.5. To achieve a similar synthesized beam for the two ammonia lines, we tapered the visibilities and applied a common restoring beam of 5.44\arcsec.
To recover the extended ammonia emission, we fill in the missing flux from the Green Bank Telescope (GBT) data. A detailed description of the GBT data reduction can be found in~\cite{sokolov+2017}.
Before merging the data, the GBT images were converted to spectral flux density units and convolved to the VLA \edit1{channel separation} of 0.2 \kms\edit1{ from the original channel separation of 0.0386 \kms}. After regridding both GBT spectral cubes to match the VLA grid, we apply the VLA mosaic primary beam response to the GBT images.
\edit1{The VLA-only images for both \nhhh inversion lines were then combined with the processed GBT data using \textsc{feather} task within the CASA package as an intermediate step, to obtain information about the spatial distribution of the extended ammonia emission. Finally}, the VLA (1,1) and (2,2) images were deconvolved again with \textsc{tclean} task, with the \textsc{tclean} mask \edit1{determined from the aforementioned combined data}. 
Constructing the clean mask from independently feathered images ensures that unbiased knowledge of the extended emission is incorporated into the \textsc{tclean} run. For specific details on the imaging strategy, we refer to Appendix~\ref{app:clean}, where a full description of the imaging and dataset combination is presented.
The resulting \nhhh (1,1) and (2,2) spectral cubes, gridded into 1\arcsec~pixels, have a common restoring beam of 5.44\arcsec.
The typical rms value of the emission free spectra in the resultant cubes is 14 $\mathrm{mJy~beam^{-1}}$ for \nhhh (1,1) and 5 $\mathrm{mJy~beam^{-1}}$ for \nhhh (2,2) inversion lines in a 0.2 \kms channel (1.04 K and 0.37 K, respectively).

\subsection{Overview of the data}

\begin{figure*}
    \makebox[\textwidth][c]{\includegraphics[width=1.35\textwidth]{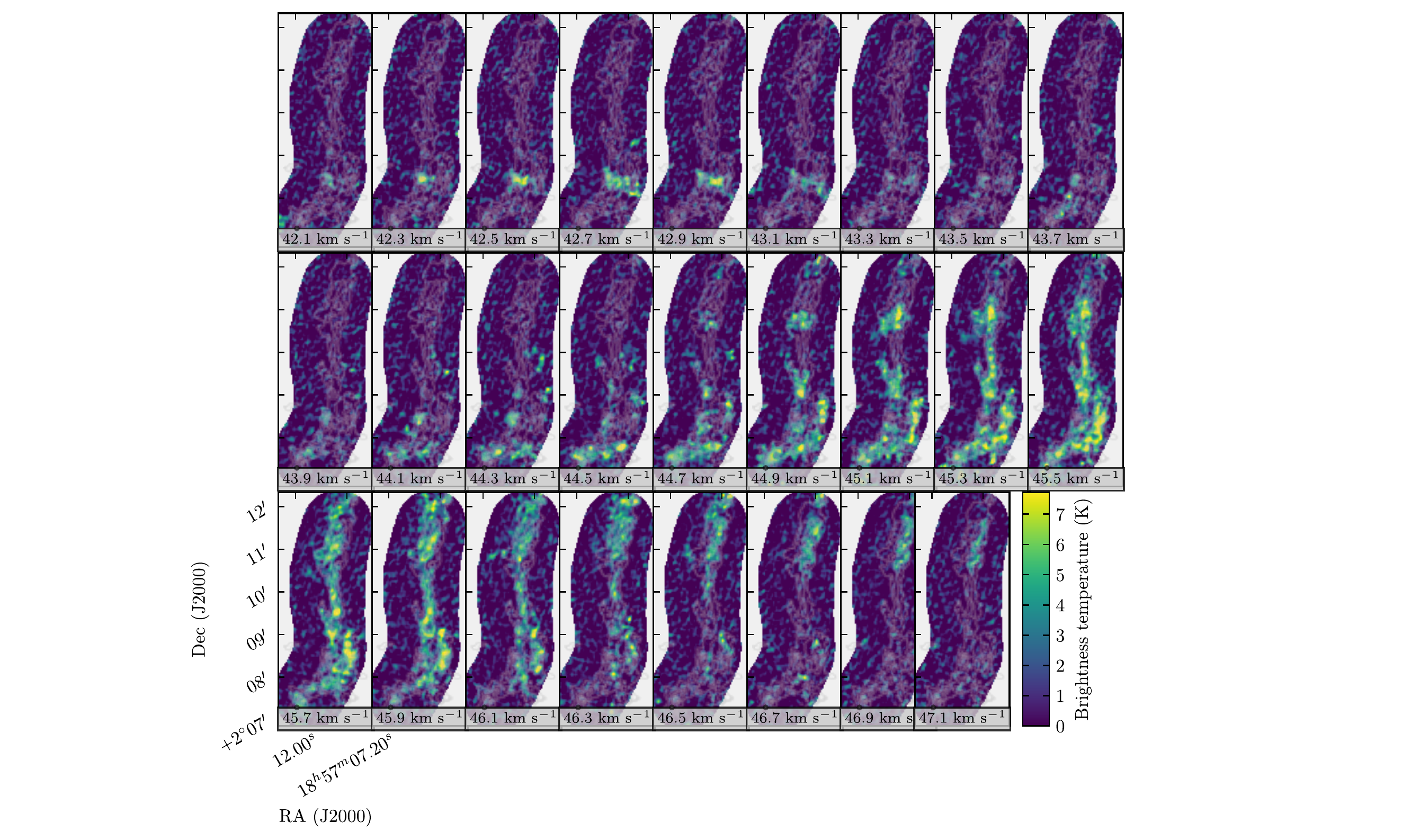}}
    \caption{Channel maps of the ammonia (1,1) line for the IRDC G035.39. Each panel shows the brightness temperature of the \nhhh (1,1) spectral cube channel of 0.2 \kms width, with text boxes at their bottom indicating the velocity centroid of the channel. The velocity range of $42\mbox{--}47$ \kms was chosen to capture the main hyperfine group of the inversion transition. The light gray contours show the infrared extinction contours starting from $A_{\rm V}=30$ mag and progressing inwards in steps of 30 mag \citep{kainulainen+tan2013}.}
    \label{fig:chanmap11}
\end{figure*}

Figure~\ref{fig:chanmap11} presents the channel maps for \nhhh (1,1) line, showing both a presence of multiple components throughout the IRDC as well as a line centroid change towards redshifted regime as one follows the IRDC northwards. The equivalent figure presenting the channel maps for the \nhhh (2,2) line is shown in Fig.~\ref{fig:chanmap22} of Appendix~\ref{app:chanmap22}, showing similar morphology.

The parsec-scale ${\sim} 0.2$ \kmspc quasi-linear velocity gradient, found in the GBT-only data~\citep{sokolov+2017} is present in our combined data%
\edit1{: the IRDC kinematics reveal a gradual change into blue-shifted velocity regime as the filament is followed southwards.} 
Superimposed on this kinematics feature, additional velocity components are present.
In the northern portion of the cloud, the western side of the IRDC splits off into a secondary, red-shifted, velocity component located around 47 \kms, in agreement with the previous studies of the cloud~\citep[labelled F3 in][]{henshaw+2013, henshaw+2014, jimenez-serra+2014}.
A filamentary-like feature connects the northern region of the IRDC with the southern one, where it appears to split into two individual filaments that then join together at the southern part of G035.39.
Additionally, the velocity feature between $42\mbox{--}43$ \kms, seen before as a clump-like structure in the ammonia maps of~\cite{sokolov+2017}, is now resolved into a distinct filament, orthogonally oriented to the rest of the IRDC.

\subsection{Line fitting}\label{ssec:fitting}
Molecular line emission tracing the dense gas within G035.39 is known to exhibit multiple line-of-sight velocity components~\cite[e.g.,][]{henshaw+2013, henshaw+2014, jimenez-serra+2014}.
However, with no clear prior knowledge of the kinematical structure exhibited in the high-resolution ammonia emission, standard algorithms commonly used to fit the line profile are prone to converge away from the global minima or, alternatively, to diverge in the absence of a proper starting point of the algorithm.
As our goal is to distinguish the physical properties of velocity-coherent structures, it is crucial to conduct a simultaneous fitting of all line-of-sight components, constraining their line parameters whenever a significant additional feature is present in the spectral profile. At the same time, overfitting has to be ruled out - that is, only statistically significant components have to be taken into consideration for the subsequent scientific analyses.

As the VLA spectral cubes consist of about a dozen thousand spectra with S/N ratio higher than three, the common practice of fitting multiple components by eye would require an excessive amount of human interaction resulting in low level of reproducibility, while methods for global non-linear regression are prohibitively computationally expensive for such a high volume of data.
Therefore, to fit multiple line-of-sight line components, we employ the same fitting strategy as in~\cite{sokolov+2017}, where an open-source Python package\footnote{\url{https://github.com/vlas-sokolov/multicube}} is used to assist the Levenberg-Marquardt fitter to find an optimal starting point for exploring the parameter space of the spectral model. The free parameters for the ammonia model~\citep[same formalism as in, e.g., ][]{friesen+pineda+2017}, implemented within \textsc{pyspeckit} package~\citep{pyspeckit}, are split into a multi-dimensional parameter grid, and a synthetic ammonia model is then computed for each. Subsequently, every spectrum in our VLA data is then cross-checked with every model computed to calculate the corresponding residual. The model parameters for a given spectrum that yield the best residual (i.e., the least sum of squared residuals) are taken to be the starting value for the Levenberg-Marquardt fitting algorithm.

Three independently-derived sets of best fitting parameters are obtained in such a fashion, for one-, two-, and three-component models\footnote{Upon close inspection, no spectra containing more than three velocity components were identified.}.
The three fitting sets are then merged into one following a set of heuristical criteria that determines the number of line-of-sight components in a given pixel of the spectral cube. The heuristical procedure is described below:

\begin{enumerate}
    \item First, the three component model is considered to be valid.
    \item If any of the three components are below signal-to-noise of three --- for both \nhhh (1,1) and (2,2) transitions --- the model is marked as rejected. Similarly, should the \edit1{half of the sum of any neighboring components' FWHM} be less than the separation between their velocity centroids (i.e., the FWHM are not allowed to overlap), the model is rejected. This restriction is imposed to avoid line blending.
    \item A rejected model is replaced with a simpler one: the three-component model is replaced with the two-component one, and a rejected two-component model is replaced with a one-component model.
    \item The replaced models are required to meet the same heuristical criteria as in step 2.
    \item If none of the fitted models are valid, we consider the spectrum to be a non-detection.
    \item The steps $1-5$ are repeated for all the spectra in the combined spectral cube, until a collection of ammonia fitting parameters is assembled in a PPV space.
\end{enumerate}

In the above procedure, the second step requires imposing constraints for both (1,1) and (2,2) lines. This is done to assure that the physical properties that depend on the population ratio of the two transitions (kinetic temperature and ammonia column density) are reliably constrained. Throughout this paper, this is the method we refer to as \textit{strict censoring} criterion for arriving at the PPV structure in the IRDC.
However, we note that in principle, should only the kinematic information be needed from our ammonia data, either one of the inversion line detections would constrain it in our simultaneous line fitting.
Therefore, we consider an additional method, which we call \textit{relaxed censoring}, which naturally results in a PPV structure with larger average number of line-of-sight components. Despite the differences in these two approaches, the average kinematic properties we will derive later will not be significantly changed if a relaxed censoring method is used (Appendix~\ref{app:acorns-kin} provides quantitative comparison between the two approaches), therefore we use the strict censoring approach because it allows us to discuss the temperature structure of the velocity components in addition to their kinematics.

\begin{figure*}
    \centering
    \includegraphics[width=.8\textwidth, trim={1cm 1.5cm 1cm 2cm}]{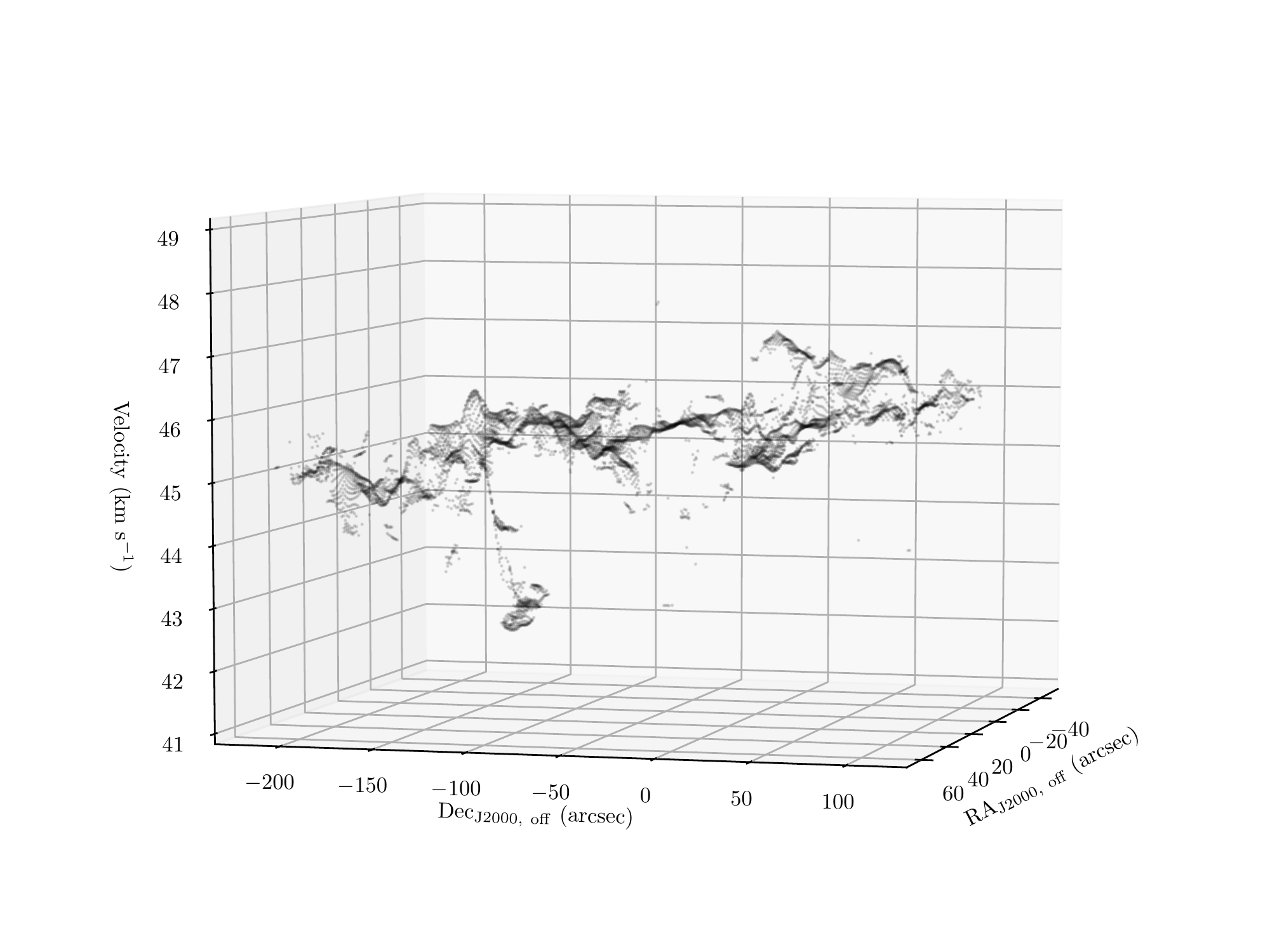}
    \caption{A PPV diagram of all fitted velocity components within IRDC G035.39. Every point in the diagram represents a best-fit velocity centroid for a single ammonia line component, and darker shade of black indicates greater density of points along the on-screen projection. When constructing this diagram, we required the signal-to-noise ratio of both (1,1) and (2,2) lines to be greater than three. The coordinates axes are specified relative to the offset at $\mathrm{\alpha(J2000) = 18h57m08s}$, $\mathrm{\delta(J2000) = +2\arcdeg10\arcmin30\arcsec}$.
    An interactive version of the diagram will be maintained under \url{https://vlas-sokolov.github.io/post/cloudh-ppv/}.}
    \label{fig:ppv}
\end{figure*}

\section{Results} \label{sec:results}

\subsection{Velocity components}\label{sec:acorns}

\begin{figure*}
    \centering
    \includegraphics[width=.95\textwidth]{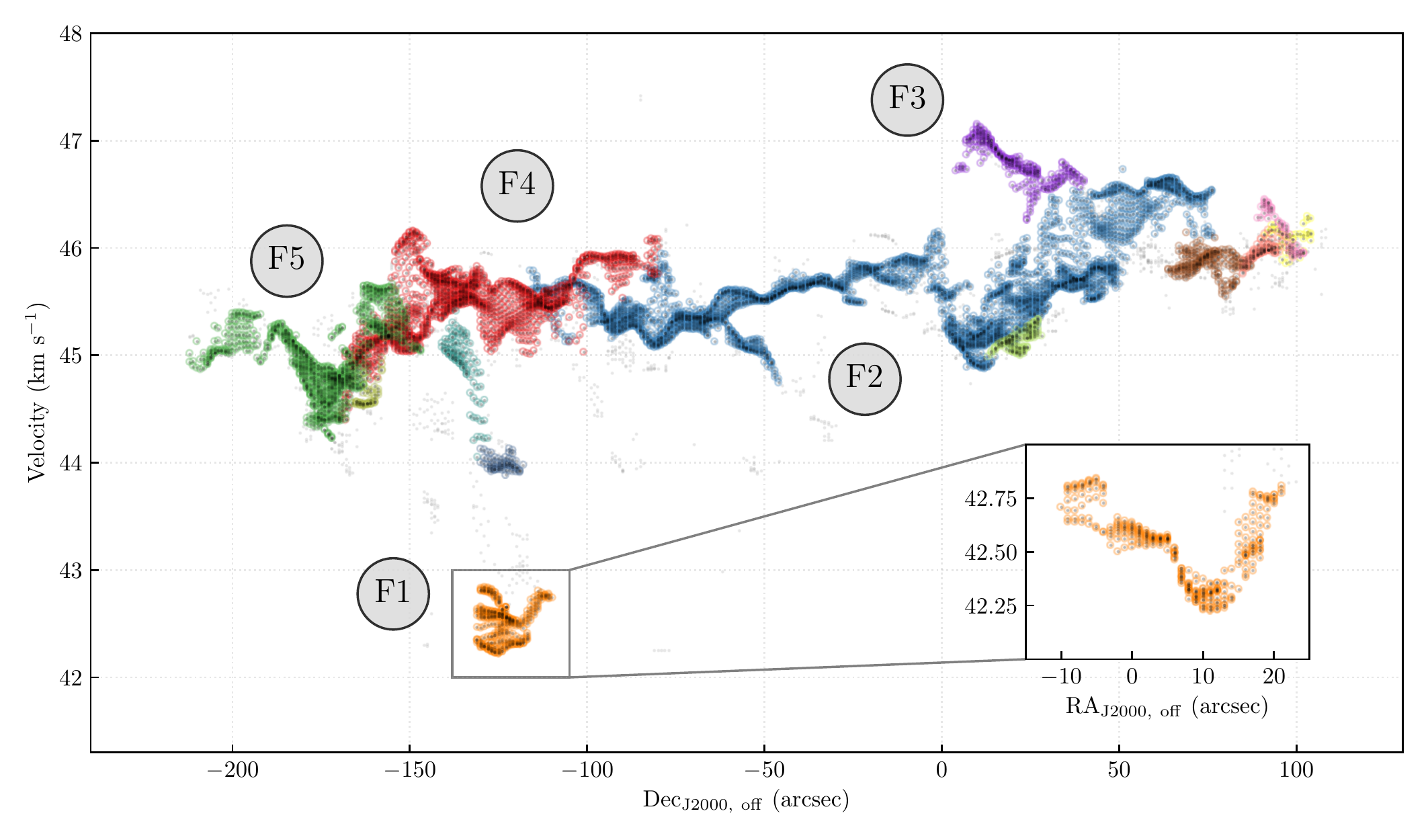}
    \caption{A PPV-diagram of the fitted velocity components within IRDC G035.39 along the Right Ascension projection. The coordinates are given in arcsecond offset relative to the $\mathrm{\alpha(J2000) = 18h57m08s}$, $\mathrm{\delta(J2000) = +2\arcdeg10\arcmin30\arcsec}$ coordinate.
All the data are plotted in black, similarly to Fig.~\ref{fig:ppv}, and individual velocity components are marked in different colors. The data not found to be associated with any clusters are plotted in gray. The figure shows the clustering obtained with the strict masking criteria (introduced in \S\ref{ssec:fitting}); the equivalent figure for relaxed censoring can be found in Appendix~\ref{app:acorns-kin}. In addition to the R.A. projection, a projection along Dec. is shown in the inset axis for the F1 filament.}
    \label{fig:acorns-dyn}
\end{figure*}

Figure~\ref{fig:ppv} shows the position-position-velocity (PPV) space with the best fit line-of-sight components, which demonstrates the complex kinematics of the G035.39 cloud.
As different parts of the cloud may represent unique, non-interacting physical structures, it is necessary to identify the components that are linked in both position and velocity. The inability to separate such unique structures would hinder quantitative analyses on the gas motions within the IRDC.

Since non-hierarchical algorithms have problems with extended emission \citep[e.g.,][]{pineda+2009} we use a hierarchical clustering of discrete PPV data from the G035.39 dataset.
For this task, we make use of the
Agglomerative Clustering for ORganising Nested Structures (ACORNS), a clustering algorithm to be fully described in Henshaw et al. (in prep.)~\citep[cf.][]{barnes+2018}. In simple terms, ACORNS links PPV points together into clusters based on a hierarchical agglomerative clustering. The algorithm starts with the most significant data point, consequently agglomerating surrounding points in the order of descending significance\footnote{For our combined ammonia data, we set the ``significance'' of the datapoints in ACORNS to be the peak signal-to-noise ratio of the combined \nhhh (1,1) and (2,2) spectra.} until a hierarchy is established.
After the first cluster has been assigned in such fashion, the most significant point among unassigned dataset is taken, and the linking is repeated. The procedure runs until no data points are left unassigned, and a dendrogram-like cluster hierarchy, assigning clusters to branches and trees~\citep[in a manner similar to][]{rosolowsky+2008-dendrograms} is established.

The parameters that defined the linking criteria of the ACORNS algorithm were the following. The clusters were restricted to be at least 9 pixels in size to avoid spurious components, and the link between neighboring datapoints was set to be smaller or equal to the velocity resolution of the data (0.2 \kms). Additionally, no links with line widths of twice the velocity resolution (0.4 \kms) were allowed to be established to prevent spatially close yet distinct components from being merged.
In our ACORNS run, the PPV points below signal-to-noise significance of three were set to be excluded.

We obtain two sets of results from the ACORNS clustering for the two masking strategies introduced in \S\ref{ssec:fitting}. The strict censoring results in thirteen hierarchical clusters, seven of which cover an area in excess of five synthesized VLA beams (i.e. have physical properties constrained in more than 116 pixels). The overall view of the structures identified is shown on Fig.~\ref{fig:acorns-dyn}, while the separated maps of the seven largest components will be presented in the following sections.
The relaxed masking results in twenty-four components, of which ten are spanning a significantly large area (a comparison with the adopted method is shown in Appendix~\ref{app:acorns-kin}).

\subsubsection{Continuum sources within the velocity components}\label{ssec:related-sources}
Throughout this study, we will relate the continuum sources discovered towards the IRDC, in particular the mass surface density cores of~\cite{butler+tan2012} and the 70~\micron sources of~\cite{nguyen_luong+2011}, to the continuous ammonia-emitting regions identified as velocity-coherent structures in PPV-space. In order to establish whether a given continuum source is related to the ACORNS-identified components, we use the following heuristics. Firstly, we require at least 24 pixels with significant ammonia spectra detections 
(corresponding to an effective area larger than that of one VLA beam)
to be enclosed in a three-beam diameter around the coordinate of interest.
Secondly, to assert the direct proximity of the continuum source to the velocity component under consideration, we require at least one significant ammonia spectra detection within one beam radius around the position of interest.
These heuristics are computed on every continuum source of interest, and for every velocity component detected.
Sources that match the criteria above will be referred to as sources related to the ACORNS component throughout this study.

Table~\ref{table:acorns-dyn} gives a summary of the velocity components we identify in this section, and lists their corresponding related sources along with the average properties of the components. The component names follow their respective colors on Fig.~\ref{fig:acorns-dyn}, and only those spanning in excess of five beam sizes are listed. In case a source appears to be related to more than one component, it is listed several times in the table.

\begin{deluxetable*}{lcccccc}
\tablecaption{The velocity components identified with ACORNS, listed by decreasing spatial extent.\label{table:acorns-dyn}}
\tablecolumns{7}
\tablewidth{0pt}
\tablehead{
    \colhead{Component ID\tablenotemark{a}} & \colhead{Size\tablenotemark{b}} & \colhead{\Vlsr\tablenotemark{b}} & \colhead{\sig\tablenotemark{b}} & \colhead{\Tkin\tablenotemark{b}} & \multicolumn{2}{c}{Related sources} \\
    \cline{6-7}
    \colhead{} & \colhead{(pixels)} & \colhead{(\kms)} & \colhead{(\kms)} & \colhead{(K)} & \colhead{\citep{butler+tan2012}} & \colhead{\citep{nguyen_luong+2011}}
}
\startdata
F2 (Blue)    & 2981 & $ 45.62 \pm 0.41 $ & $ 0.28 \pm 0.14 $ & $ 12.29 \pm 2.18 $ & H5, H6  & \#12, \#28, \#6, \#20, \#5 \\ 
F4 (Red)     & 1377 & $ 45.46 \pm 0.30 $ & $ 0.31 \pm 0.12 $ & $ 11.69 \pm 1.70 $ & H2, H4  & \#22, \#10, \#16           \\ 
F5 (Green)   & 927  & $ 44.96 \pm 0.32 $ & $ 0.32 \pm 0.14 $ & $ 12.40 \pm 2.03 $ & H1      & \#7                        \\ 
F1 (Orange)  & 313  & $ 42.52 \pm 0.17 $ & $ 0.17 \pm 0.06 $ & $ 11.22 \pm 1.29 $ & H3      & \#11                       \\ 
F6 (Brown)   & 198  & $ 45.86 \pm 0.10 $ & $ 0.33 \pm 0.25 $ & $ 11.20 \pm 1.76 $ & -       & -                          \\ 
F3 (Purple)  & 270  & $ 46.79 \pm 0.18 $ & $ 0.24 \pm 0.06 $ & $ 12.12 \pm 1.90 $ & -       & \#18                       \\ 
F7 (Olive)   & 124  & $ 45.14 \pm 0.09 $ & $ 0.15 \pm 0.05 $ & $ 13.47 \pm 1.94 $ & H6      & \#6                        \\ 
\enddata
\tablenotetext{a}{The color labels directly correspond to the component colors on Fig.~\ref{fig:acorns-dyn}.}
\tablenotetext{b}{The mean followed by the standard deviation of the values belonging to the given component.}
\end{deluxetable*}

\subsection{Velocity structure across the IRDC}\label{ssec:velgrad}

Various processes in star-forming clouds, such as, turbulence, gravitational acceleration of infalling material, feedback from the embedded sources, or rotation can create line-of-sight velocity gradients.
In order to perform a qualitative analysis of the velocity gradients, we have to interpolate the slope of discrete \Vlsr measurements on the plane of sky. To achieve this, we use a variation of the Moving Least Squares method~\citep{lancaster+salkauskas1981} for constructing continuous surface from a sample of discrete data. The method, often used in computational geometry and computer vision~\citep[e.g.,][]{levin1998, schaefer+2006}, was independently developed
in astrophysics by~\cite{goodman+1993} to find velocity gradients from line of sight velocity fields.
The automated local velocity gradient fit, which applies the \citeauthor{goodman+1993} method across the cloud to study the internal velocity field, was developed by \cite{caselli+2002-n2hp}, and has seen considerable usage since~\citep[e.g.,][]{barranco+goodman1998, caselli+2002-l1544, tafalla+2004, kirk+2013, henshaw+2014, henshaw+2016_scouse}.

The local line-of-sight velocity gradient is fit to a 2D scalar field given by the nearby PPV points, approximated by a first-degree bivariate polynomial $ f(\alpha, \delta) = v + a \Delta \alpha + b \Delta \delta $, where the R.A. and Dec. offsets $\Delta \alpha$ and $\Delta \delta$ are confined to a circle of radius $R$ around a given set of sky coordinates $\mathbf{r} = (\alpha, \delta)$, and $(v, a, b)$ are the free parameters of the least squares fit ($v$ is the centroid velocity and $\sqrt{a^2 + b^2}$ is the velocity gradient magnitude). For each velocity component identified and for every position $\mathbf{r}$, we construct a set of valid pixels as follows. First, only the pixels with a well defined \Vlsr value (i.e. the relevant spectral component having a peak $\mathrm{S/N} > 3$) are included in the velocity gradient analysis, and the rest are masked. Secondly, the moving least squares operates only on the values within three-beam diameter (i.e., $R = 1.5 \times 5.44\arcsec$) and only if more than 46 centroid velocity measurements (i.e. filling an effective area larger than two VLA beams) are defined within it. Finally, only spatially continuous regions are considered, and \Vlsr regions within radius $R$ that are not connected with position $\mathbf{r}$ with valid \Vlsr pixels are masked as well. For every valid point with fitted \Vlsr value, an optimal set of parameters $\mathbf{p}(\mathbf{r}) = a_{\mathbf{r}}, b_{\mathbf{r}}$ is obtained by a weighted least-squares fit to a collection of nearby pixels:
$$ \mathbf{p}(\mathbf{r}) = \operatorname*{argmin}_{a, b} \left\{\sum_{\mathbf{r'}: \| \mathbf{r} - \mathbf{r'} \| \leq R} \frac{(v_{\mathrm{lsr}}(\mathbf{r'}) - f(\mathbf{r'}, a, b))^2}{w(\mathbf{r'})^2}\right\}, $$

\begin{figure*}
    \centering
    \includegraphics{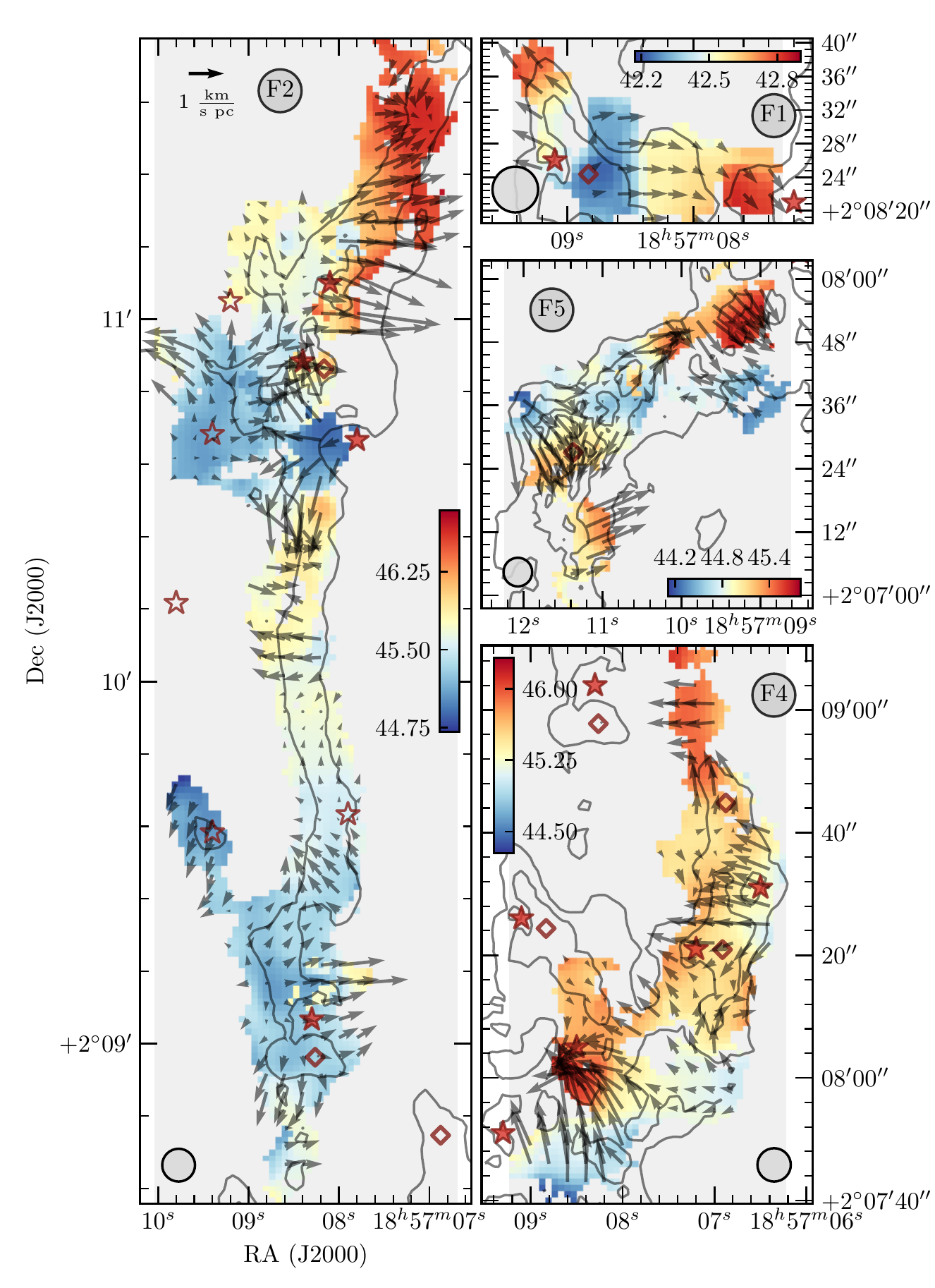}
    \caption{A map of derived $V_{\mathrm{lsr}}$ values for the biggest velocity components (left panel: F2; right panels, top to bottom: F1, F5, F4), overlaid with the velocity gradient arrows. The directions of the arrows points to the steepest velocity field change in the red-shifted direction, while the arrow lengths represent the relative vector magnitudes of the gradient. The open and filled red stars denote the positions of 70~\micron Herschel sources from~\cite{nguyen_luong+2011} below and above 20 \Ms, respectively, and the red diamonds indicate the position of cores from~\cite{butler+tan2009, butler+tan2012}. The overlaid contours indicate the highest extinction contours from~\cite{kainulainen+tan2013}, ranging from $\mathrm{A_V}=40$ to $120$ mag, progressing in steps of 20 mag. \edit1{The relative position on the plane of sky of the components shown is displayed in App. \ref{app:relcomp}.}}
    \label{fig:velgrad}
\end{figure*}

\noindent where $\mathbf{r'} = (\alpha + \Delta \alpha, \delta + \Delta \delta$) is a vector on the plane of sky, and $w(\mathbf{r'})$, the weights of the fit at a given position, were set to be the fitting uncertainties on the \Vlsr. 

Following~\cite{goodman+1993}, we derive the velocity gradient magnitude \velgradmag and orientation $\theta$ for all the velocity components. Additionally, we propagate each pair of best-fit $(a, b)$ values into uncertainties on \velgradmag and $\theta$. 
%
%
%
The dominant majority (96\%) of the values derived are statistically significant ($\mathrm{S/N} > 3$), with a mean value of uncertainties on \velgradmag of 0.06 \kmspc. Fig.~\ref{fig:velgrad} shows the \Vlsr for the four largest velocity components (F2, F4, F5, and F1), overlaid with the velocity gradients derived for them.

\begin{figure}
    \centering
    \includegraphics[width=.5\textwidth]{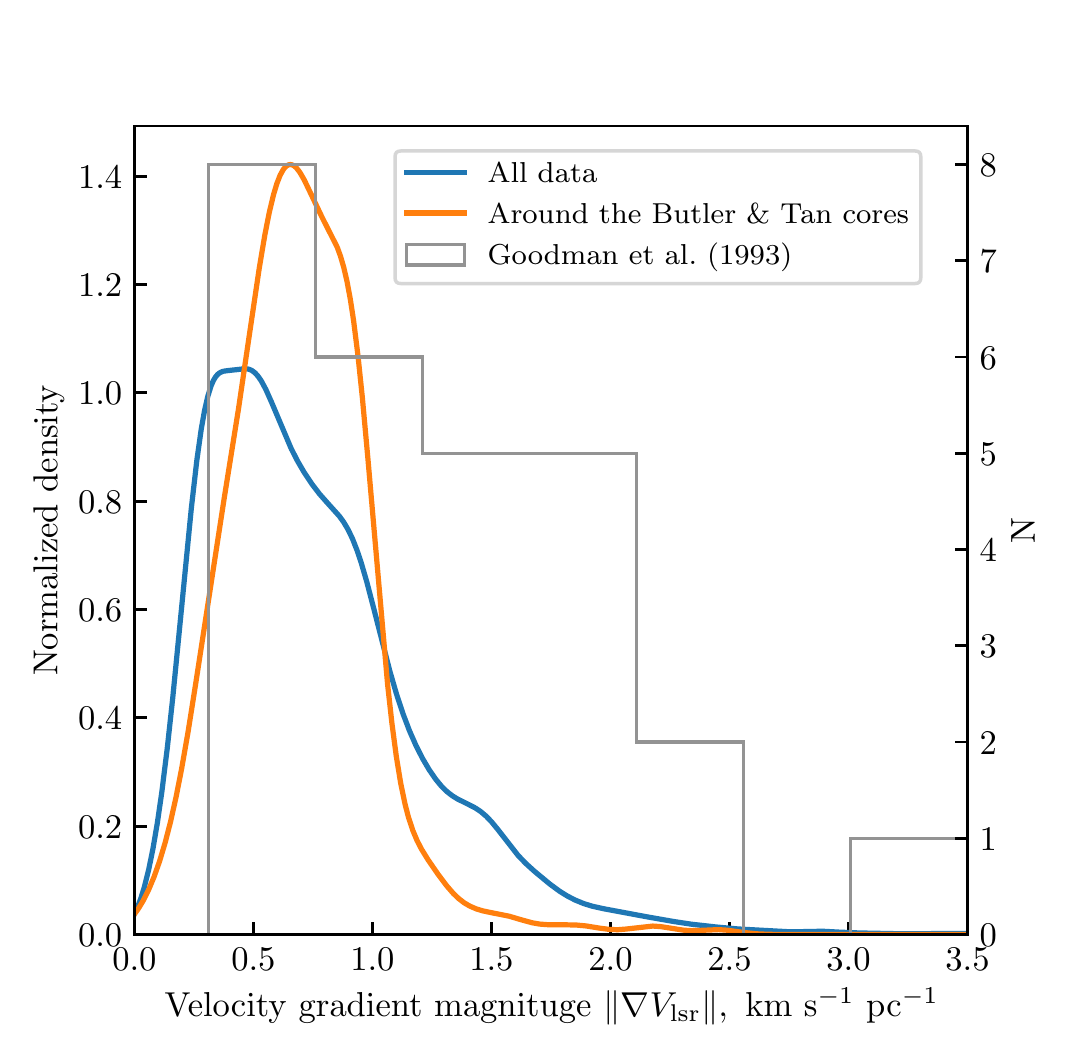}
    \caption{Kernel density estimate (KDE) of the velocity gradient magnitudes for all the velocity components derived. In addition to the total distribution of \velgradmag shown in blue, values within three beam diameter separation from~\cite{butler+tan2009, butler+tan2012} cores (following the selection criteria outlined in Section~\ref{ssec:related-sources}) are shown in orange. The KDE bandwidth selection was performed following the Scott's Rule~\citep{scott1992}. A histogram of statistically significant values ($\mathrm{S/N} > 3$) of velocity gradient magnitudes in low-mass dense cores is shown for comparison~\citep[adapted from][]{goodman+1993}.}
    \label{fig:gradhist}
\end{figure}

The overall velocity gradient magnitude distribution peaks at 0.35 \kmspc (Fig.~\ref{fig:gradhist}, values with ${\rm S}/{\rm N} < 3$ are not shown) and shows an asymmetrical tail towards larger values.
To study the potential effect of continuum cores dynamically impacting the magnitude of large-scale dense gas motions in the IRDC, we have selected a sub-sample of the velocity gradient magnitudes that we consider to be spatially relevant to the~\cite{butler+tan2009, butler+tan2012} cores. While the distribution of the core-adjacent \velgradmag values appears to be narrower than the overall distribution, possibly due to smaller dynamic ranges of density probed, the core-adjacent values are representative for the whole cloud with no statistically significant enhancements.

When compared to studies of nearby low-mass dense cores probing similar spatial scales, the gradient magnitude values we find are consistently smaller than those of~\cite{caselli+2002-n2hp}, who have performed a similar velocity gradient analysis on a large sample of low-mass dense cores with a dense gas tracer \nnhp, finding typical velocity gradients of $0.8-2.3$ \kmspc.
The difference can be attributed to the higher spatial resolution used in the study (corresponding to about 0.04 pc, calculated by taking a median distance to the core sample).
A closer comparison can be made with the~\cite{goodman+1993} results, which are obtained both with the same molecular transition, \nhhh (1,1) as well as with similar physical scales probed (0.07 pc from the median distance to \citeauthor{goodman+1993} core sample), yielding the velocity gradient magnitudes of $0.3-2.5$ \kmspc, which is broadly consistent with our findings in G035.39 (Fig.~\ref{fig:gradhist}). 

\subsection{Temperature maps}

\begin{figure}
    \centering
    \includegraphics[width=.46\textwidth]{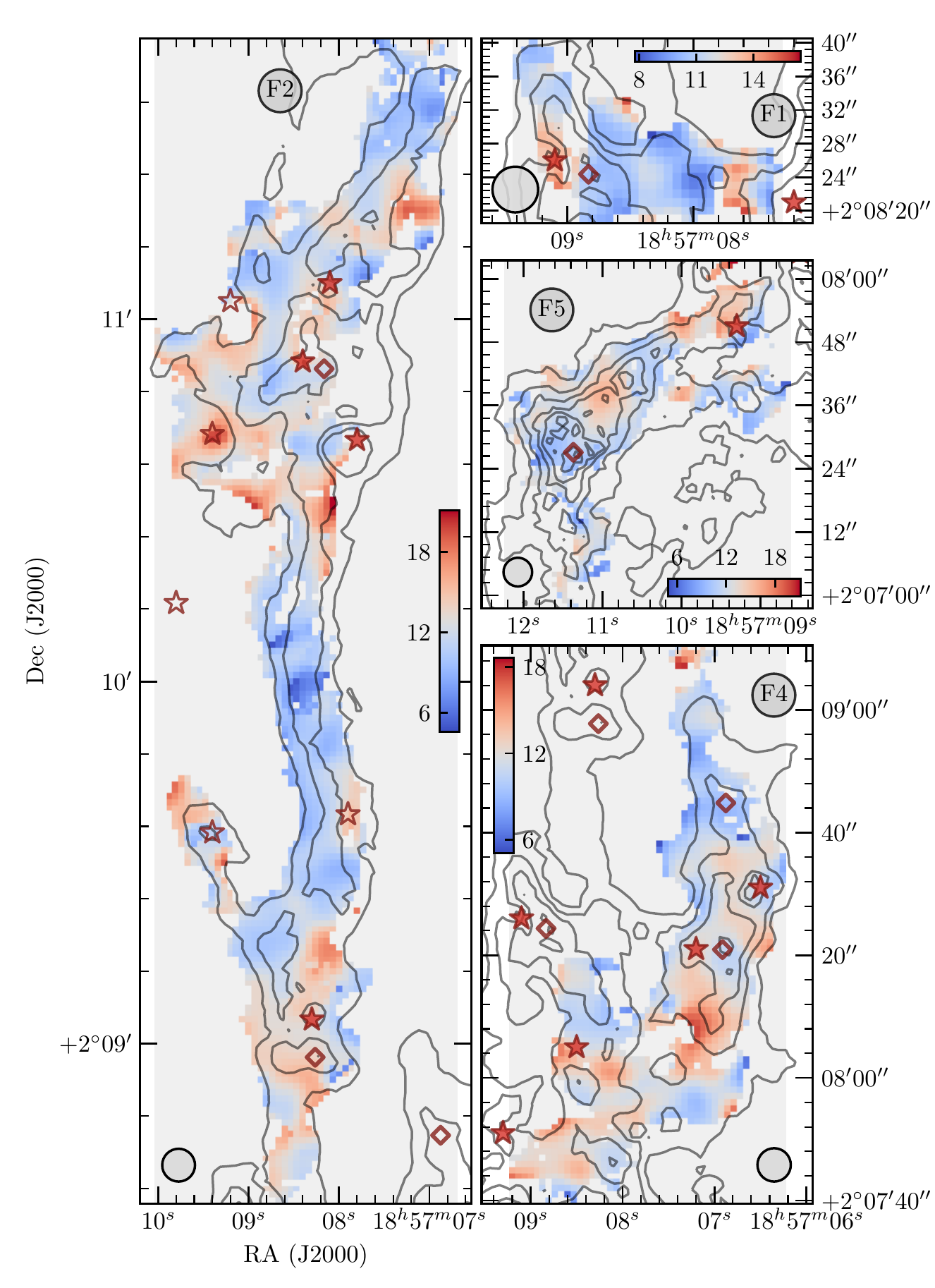}
    \caption{A map of derived kinetic temperature values for the largest velocity components. The colorbar units for \Tkin are in Kelvin. The layout and the source markings match those on Fig.~\ref{fig:velgrad}. The overlaid contours show the extinction contours from~\cite{kainulainen+tan2013}, ranging from $\mathrm{A_V}=30$ to $120$ mag, progressing in steps of 15 mag. \edit1{The relative position on the plane of sky of the components shown is displayed in App. \ref{app:relcomp}.}}
    \label{fig:tkin}
\end{figure}

Following the line fitting and component separation procedures outlined in \S\ref{ssec:fitting} and \S\ref{sec:acorns}, respectively, we arrive at the highest angular resolution temperature maps for G035.39. Furthermore, the sensitivity of our combined observations allows us to resolve the temperatures for the line-of-sight components in the IRDC.
The spatial distribution of the ammonia-derived gas kinetic temperature values is presented in Figure~\ref{fig:tkin} for the four largest IRDC components (F2, F4, F5, and F1). The mean of the overall temperatures for all the spectra analyzed in G035.39 is 11.8 K, with the standard deviation of 2.9 K, and the corresponding statistics for individual components are listed in Table~\ref{table:acorns-dyn}.

\begin{figure*}
    \centering
    \hspace*{1.1cm}\includegraphics[width=.75\textwidth]{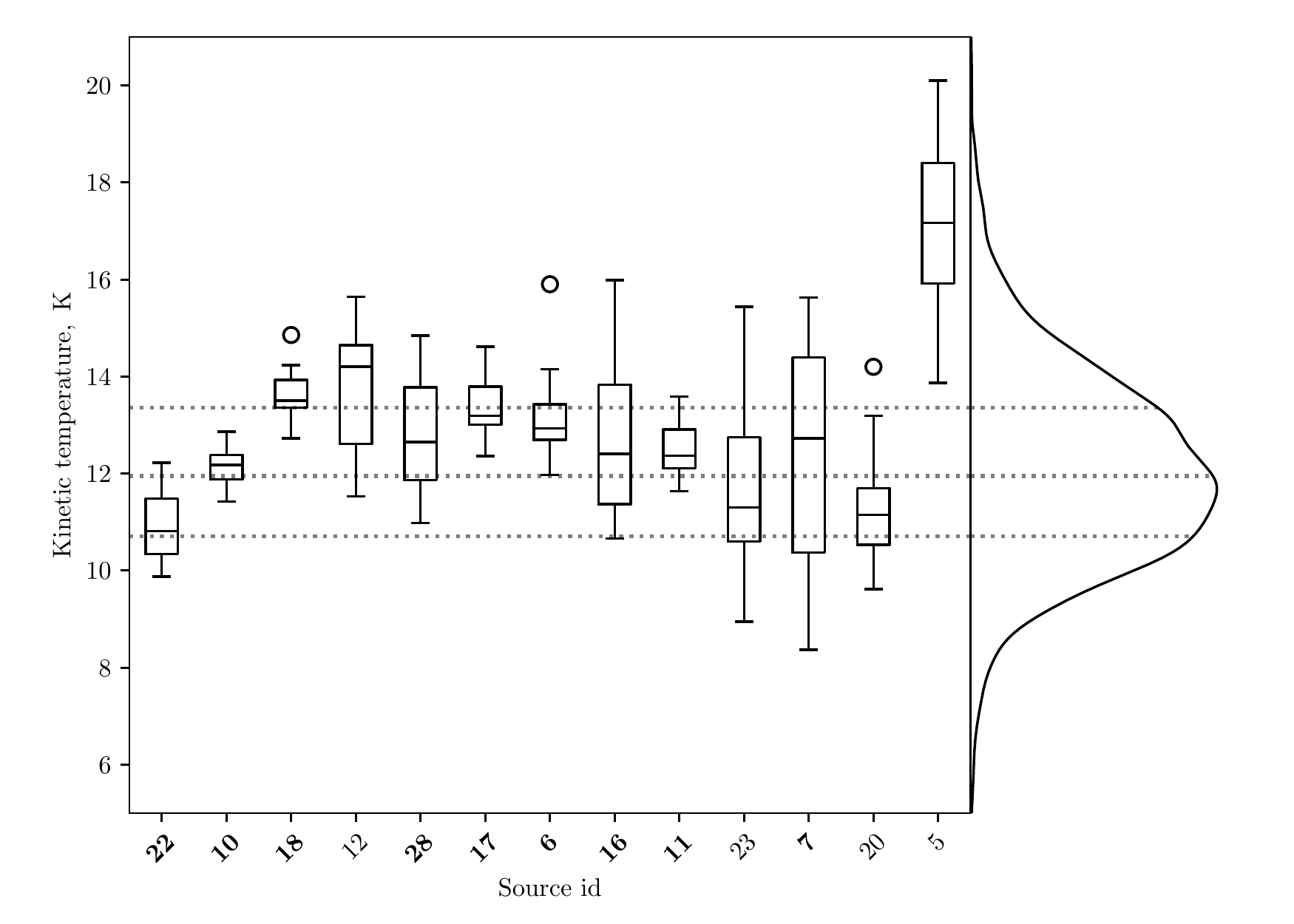}\hspace*{-1.1cm}
    \caption{The box plots show the systematic temperature enhancements seen within one VLA beam around each of the 70~\micron sources (following the selection criteria outlined in Section~\ref{ssec:related-sources}).
    The source number follows \cite{nguyen_luong+2011}, with the low-mass dense cores labeled alongside the massive ($M > 20$ \Msol) dense cores in bold.
    The overall distribution (KDE, solid line) of the kinetic temperature values for all the velocity components derived, and its 25$^{\rm th}$, 50$^{\rm th}$, and 75$^{\rm th}$ percentiles (dotted lines) are shown on the right-hand side of the Figure. The KDE bandwidth selection was performed following the Scott's Rule~\citep{scott1992}.}
    \label{fig:temphist}
\end{figure*}

In the GBT-only analysis,~\cite{sokolov+2017} identified localized heating at the location of massive protostellar sources. This effect, however, was only seen as a collective contribution due to limited angular resolution of their temperature map. Here, the increased angular resolution and reliable detection of both ammonia inversion lines enables a discussion on the individual contribution of the 70~\um sources to the temperature enhancements seen in the IRDC.
Fig.~\ref{fig:temphist} shows a comparison between the \Tkin distribution and the temperature values within one VLA beam around individual 70~\micron sources.
Compared to the overall distribution of the kinetic temperature, which peaks at 11.7 K and has typical values in the $9 - 15$ K range, the \Tkin values in the vicinity of the 70~\micron sources show slight temperature enhancements. 
Temperatures consistently lower than the mean (by about 2 K) are found towards the neck-like structure connecting the north and the south parts of the IRDC (left subfigure on Fig.~\ref{fig:tkin}), which is largely devoid of protostellar sources, with the exception of one 70~\micron source on the western edge of the filament, previously classified as low-mass dense core~\citep[source \#12 in][]{nguyen_luong+2011}. The source shows moderate temperature enhancements around it, peaking at about 15 K relative to the overall mean of the filament of 10.5 K. The excellent correspondence of the position of the 70~\micron source and the localized temperature enhancement in an otherwise cold filament shows that the protostellar source is indeed embedded in the filament, ruling out the possibility of the source being a line-of-sight projection on it.

The protostellar source in the north-eastern part of the IRDC (F2 component, left panel on Fig.~\ref{fig:tkin}), classified as a low-mass dense core~\citep[$9 \pm 1$ \Msol,][]{nguyen_luong+2011}, is one of the warmest regions traced with our temperature maps.
It is one of the most active among the protostellar sources in the IRDC, manifesting 8-, 24-, and 70-\micron point-source emission.
The gas kinetic temperature increases by 8 K in its vicinity, with the peak of 20 K coinciding with the location of the 70~\um point source emission (Fig.~\ref{fig:tkin}).
With the velocity centroid of the ammonia line of 45.1 \kms continuously morphing with the kinematics of the rest of the cloud (Fig.~\ref{fig:velgrad}), the corresponding temperature enhancement is an argument in favor of the source being embedded in the dense filament traced with our ammonia observations.
The puzzling nature of the source is emphasized by the ammonia line widths found towards it. As shown in~\cite{sokolov+2018}, it coincides spatially with an extended region of subsonic turbulence, and features some of the narrowest non-thermal velocity dispersions in the whole dataset.

The velocity feature at 42.5 \kms, labelled as F1 filament in our study (expanded into a subfigure in Fig.~\ref{fig:acorns-dyn}), forms a discontinuity in PPV space with respect to the rest of the IRDC kinematics, despite appearing to trace the southernmost end of Filament 1 of \citep{jimenez-serra+2014}. This blue-shifted component shows a velocity swing of about 0.5 \kms towards its center, where an extinction core~\citep{butler+tan2009, butler+tan2012} is located, and a 70~\micron source is found next to it~\citep{nguyen_luong+2011}.
The angular separation between the 70~\micron source and the~\cite{butler+tan2009, butler+tan2012} extinction core is 4.3\arcsec, which is smaller than both the beam of our ammonia observations as well as the 70~\um~band of the SPIRE instrument. The two sources are thus likely to be related.
The 70~\um source appears as a temperature enhancement in the temperature map, which suggests that star formation is ongoing in the F1 component.

\begin{figure}
    \centering
    \includegraphics[width=.5\textwidth]{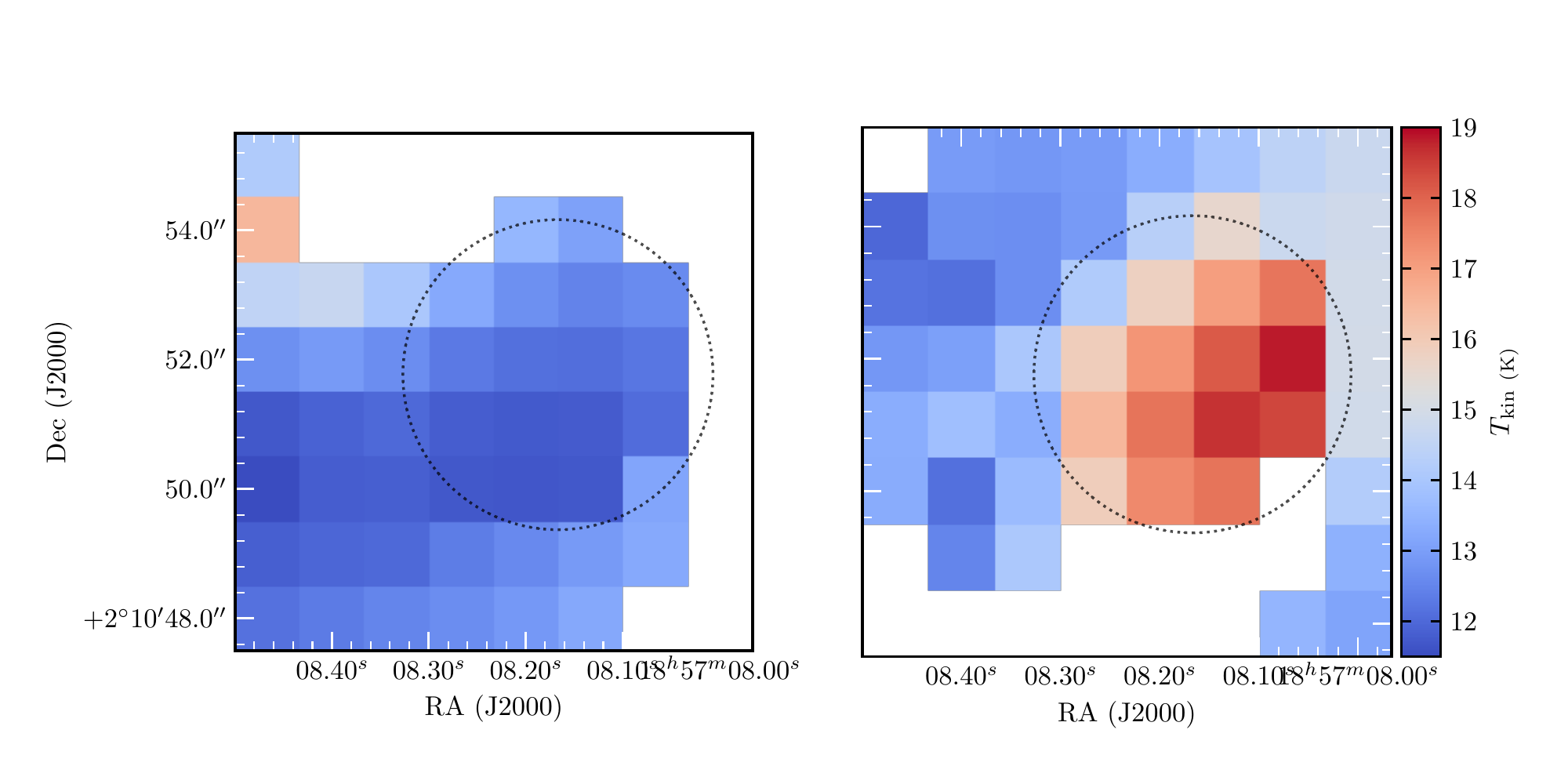}
    \caption{A zoom-in on the H6 core location for two overlapping line-of-sight velocity components, F2 (left panel) and F7 (right panel), showing derived kinetic temperatures. The black circle marks the fitted size of H6 core in~\citep{butler+tan2012}.}
    \label{fig:tkinH6}
\end{figure}

We find the H6 core in the north of the cloud to be associated with two ACORNS components, F2 and F7, due to the VLA observations resolving two continuous overlapping line-of-sight regions.
We report a 7 K increase in gas kinetic temperature in the F7 component, coinciding with the extinction peak of~\cite{butler+tan2012} and of similar size as the H6 core (Fig.~\ref{fig:tkinH6}).
The excellent correspondence between the temperature enhancement in the F7 components and the location of the core allows us to pinpoint the H6 location in the PPV space. The exact physical nature of the temperature enhancement is less certain, and we suggest that heating from embedded protostars, either as direct radiation or in a form of unresolved outflows might stand behind the heating. Despite the temperature enhancement seen in the F7 component, we do not observe significant broadening of the line at the H6 position, and the line width, in fact, points to subsonic motions as will be seen in the next section.

In G035.39, the point-like 70~\micron emission, signifying protostellar activity, is found in all but one among the~\cite{butler+tan2012} cores. Located at the peak of the extinction map in the south of the F5 filament, the H1 core appears as a cold feature in the VLA temperature map (Fig.~\ref{fig:tkin}), with a mean kinetic temperature of 10.7 K within the VLA beam, and has a complex kinematic morphology (Fig.~\ref{fig:velgrad}).
Recent observations by~\cite{liu+2018} find evidence for a pinched magnetic field morphology towards the core, indicating that the line of sight velocity gradients on~Fig.~\ref{fig:velgrad} could be an argument for accretion flows along the filament.
The mean kinetic temperature of the H1 core agrees well with that of the low-mass dense cores \citep[e.g., 11 K for a sample of 193 cores in][]{rosolowsky+2008-nh3}.

\section{Discussion}\label{sec:discussion}

\subsection{Subsonic line widths in G035.39}

Turbulence, dominating molecular clouds on large scales ($> 1$ pc), is found to dissipate towards smaller scales~\citep{larson1981}, allowing for bound prestellar core formation, and, ultimately, for star formation.
Turbulence dissipation has been detected via high-J rotational transitions of CO in quiescent clouds and IRDCs \citep{pon+2014, pon+2015, pon+2016}.
Subsonic turbulence is observed throughout low-mass star-forming regions, and is a natural consequence of turbulent energy dissipating towards smaller scales~\citep{kolmogorov1941}. Observationally, it is seen both as a transition to coherence on small scales (within 0.1 pc) as coherent cores~\citep[e.g.,][]{pineda+2010, pineda+2011}, as well as large, filament-scale subsonic regions in Taurus, Musca, Serpens South~\citep{tafalla+hacar2015, hacar+2016_lwidth, friesen+2016}.

Massive star formation on the other hand, is thought to be governed by non-thermal motions. However, in IRDC G035.39 \cite{sokolov+2018} find that substantial regions of the cloud are predominantly subsonic (i.e. with sonic Mach number $\mathcal{M} \le 1$). While the implications for the turbulence dissipation in other IRDCs are discussed in \cite{sokolov+2018}, we cite the fraction of subsonic spectra identified in the ACORNS velocity components below.
Table~\ref{table:acorns-machs} lists the average Mach numbers, the spread of their distribution (one standard deviation), and the overall fraction of subsonic components relative to the total in each of the velocity components spanning in excess of five VLA beams. In addition, details of the Mach numbers derivation, as well as the propagation of uncertainty on $\mathcal{M}$ matching that of \cite{sokolov+2018} are included as an appendix item in Appendix~\ref{app:mach}.

\begin{deluxetable}{lcc}
\tablecaption{Subsonic motions within the G035.39 velocity components.\label{table:acorns-machs}}
\tablecolumns{3}
\tablewidth{0pt}
\tablehead{
    \colhead{Component ID\tablenotemark{a}} & \colhead{$\mathcal{M}$\tablenotemark{b}} & \colhead{$\mathcal{M} \le 1$ fraction (\%)}}
\startdata
F2 (Blue)   & $1.24 \pm 0.70$ & 44 \\ 
F4 (Red)    & $1.41 \pm 0.62$ & 32 \\ 
F5 (Green)  & $1.42 \pm 0.67$ & 30 \\ 
F1 (Orange) & $0.74 \pm 0.30$ & 81 \\ 
F6 (Brown)  & $1.54 \pm 1.41$ & 55 \\ 
F3 (Purple) & $1.00 \pm 0.32$ & 48 \\ 
F7 (Olive)  & $0.54 \pm 0.25$ & 98 \\ 
\enddata
\tablenotetext{a}{The color labels directly correspond to the component colors on Fig.~\ref{fig:acorns-dyn}.}
\tablenotetext{b}{The mean followed by the standard deviation of the values belonging to the given component.}
\end{deluxetable}

We report no significant trends in the Mach numbers against temperature or density (either column density or the extinction-based mass surface density). However, we note that in two velocity components that are associated with star formation based on both the presence of 70~\um emission and kinetic temperature increase (F1 and F7; Figures~\ref{fig:tkin},~\ref{fig:tkinH6}) we observe a larger fraction of subsonic motions than in any other component (Table~\ref{table:acorns-machs}).
We suggest that future JVLA observations are needed to fully resolve the boundary over which transition to coherence in the IRDC occurs, its relation to star formation, and the degree of similarity to the transition to coherence in low-mass star-forming clouds \citep[e.g.,][]{goodman+1998, pineda+2010}, which could further probe smaller scale fragmentation \citep{pineda+2015}.

\subsection{Comparison with \nnhp}

\begin{figure}
    \centering
    \includegraphics[width=.5\textwidth]{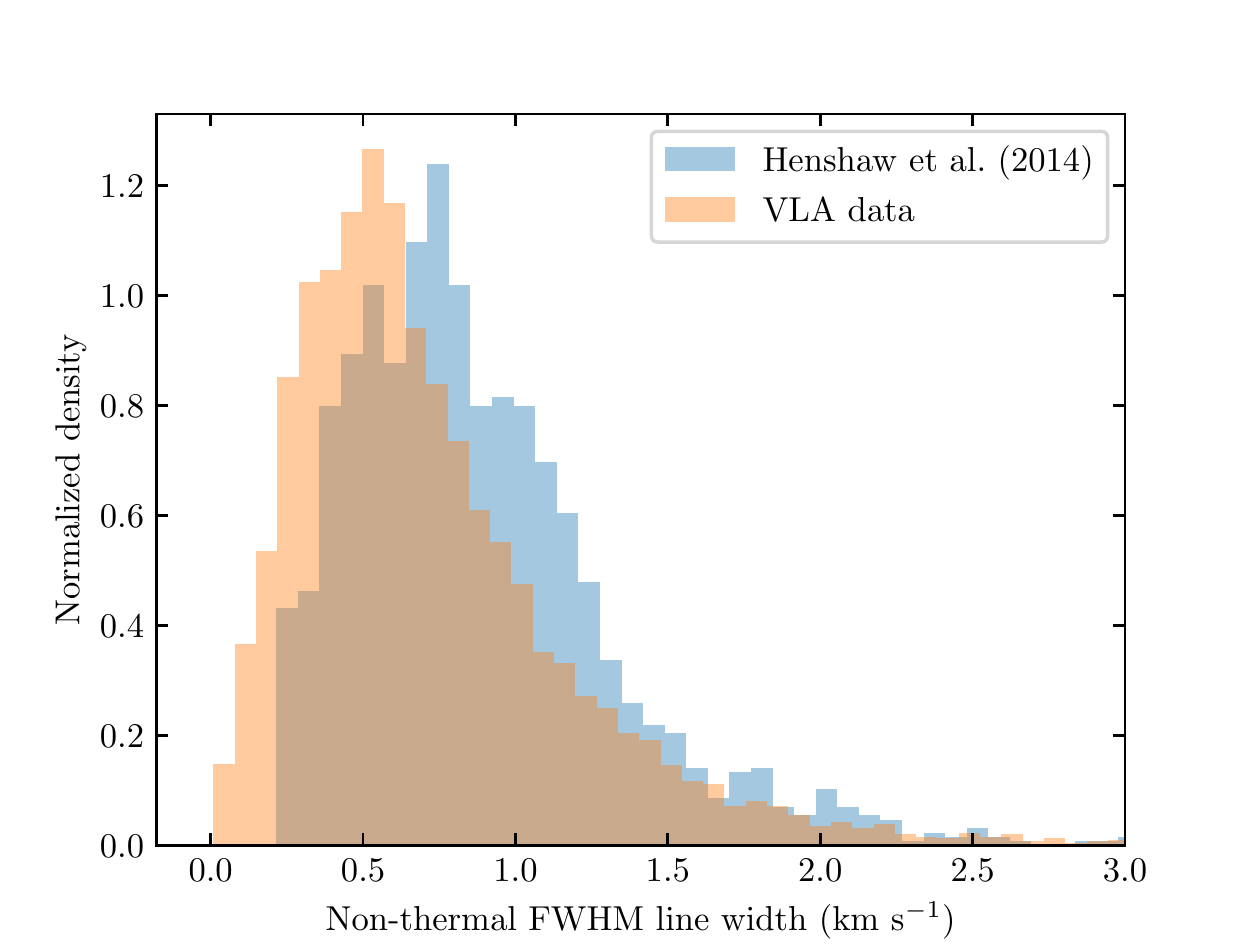}
    \caption{A comparison between the FWHM line widths derived in~\cite{henshaw+2014} and those derived in our work. As the VLA observation mosaic covers a larger area, only the values overlapping with~\cite{henshaw+2014} coverage are included.}
    \label{fig:lwidthcomp}
\end{figure}

\edit1{The region of G035.39} mapped by~\edit1{\cite{henshaw+2014}} with their Plateau de Bure Interferometer (PdBI) observations of \nnhp (1-0) line transition is fully covered in our VLA mosaic. \edit1{As the spectral setups of both datasets were consistent (final channel separation of 0.2 \kms and 0.14 \kms for VLA and PdBI data, respectively, with no extra smoothing applied)}, we are well-positioned to perform a quantitative comparison of the two tracers of dense gas kinematics.
Assuming that the \nhhh and \nnhp molecules are tracing similar gas, the turbulent (or non-thermal) components of their line widths should be equal. Figure~\ref{fig:lwidthcomp} shows a direct comparison between the two non-thermal line widths. As the \nnhp molecule is heavier than \nhhh, we have subtracted the thermal component from both transitions in making the figure for a more direct comparison, taking \Tkin from the mean ammonia kinetic temperature for both molecules to assure the uniform comparison between the two datasets. \edit1{In other aspects, the derivation of the non-thermal \nnhp line width follows our calculations for the ammonia line widths (App. \ref{app:mach}), with substitutions made for molecular mass where appropriate.} Only the data from the commonly mapped region are shown on Fig.~\ref{fig:lwidthcomp}\edit1{, regardless of the number of line-of-sight components found in either dataset.}

Surprisingly, we find the non-thermal line widths derived from our VLA observations to be about $20\%$ narrower than those of the PdBI \nnhp (1-0) from~\cite{henshaw+2014}.
This is unexpected, as~\cite{henshaw+2014} data resolution are comparable to our common VLA restoring beam size of 5.44\arcsec.
We also note that both datasets have also had the missing flux filled in from complementary single-dish observations.
In addition, the critical densities of the \nhhh (1,1) and (2,2) transitions are an order of magnitude lower than that of \nnhp (1-0)~\citep{shirley2015}, so if the higher-density gas is less turbulent, a broader line width should be seen in the ammonia spectra, contrary to our results.
However, ammonia can selectively trace high-density regions,
as its numerous hyperfine components allow to trace gas also at densities significantly larger than the critical density and its abundances increases toward the core centers \citep{tafalla+2002, crapsi+2007, caselli+2017}; thus, \nhhh is expected to follow material traced by \nnhp (1-0) as found in low-mass star forming regions \citep{benson+caselli+1998}.
Finally, the presence of magnetic fields is expected to affect the line widths of the ion species, trapping the ion molecules and narrowing their line profiles~\citep{houde+2000a}, a trend opposite to what we find here. However, the phenomenon requires large turbulent flows to occur in directions misaligned with respect to the magnetic field lines, and it was not found to come into play in dark clouds primarily supported by thermal motions \citep{houde+2000b}.

Two possible explanations of the line width discrepancy above are differences in hyperfine structure modelling and varying chemical differentiation.
In a detailed study of two starless cores,~\cite{tafalla+2004} have found a similar discrepancy in the line widths of \nhhh and \nnhp, and have attributed the differences in non-thermal line widths to non-LTE effects of the \nnhp lines~\citep[cf.][]{caselli+1995}.
Because the PdBI data from~\cite{henshaw+2014} were fit with an isolated hyperfine component only, modelling it as a simple Gaussian component with $\tau \ll 1$, the relative broadening of \nnhp line with respect to ammonia could be an optical depth effect. However,~\cite{henshaw+2014} discuss this scenario and make several arguments in favor of the isolated hyperfine transition being optically thin.
Another scenario is a potential chemical differentiation between \nnhp and \nhhh across the IRDC, preferentially tracing regions of lower turbulence in ammonia emission.
Large differences in abundance profiles of different molecules are commonly found towards low-mass starless cores~\citep[e.g.,][]{tafalla+2002, spezzano+2017} and infrared dark clouds~\citep{feng+2016}.
For the particular case of \nhhh and \nnhp, several observational works have proposed that their abundance ratio could be an evolutionary indicator in dense cold cores \citep{palau+2007, fontani+2012a, fontani+2012b}.
Chemical differentiation could then play a dominant role in setting the line width differences, outweighing the other arguments listed above that would favor a broader ammonia line width.
This, however, would require a significant difference in the abundance profiles for the two molecules, a difference that can be quantified by comparing the spatial distribution in the integrated intensities of \nhhh and \nnhp. However, both tracers appear to follow the dense gas closely, showing strong linear correlations with mass surface density from~\cite{kainulainen+tan2013}: for \nhhh (1,1) (Pearson's $r=0.73$), \nhhh (2,2) ($r=0.71$), and \nnhp (1-0)~\citep[$r \sim 0.7$,][]{henshaw+2014}. Therefore, the origin for the discrepancy remains unclear, and it is likely that higher angular resolution observations are needed to understand it.

\subsection{Complex gas motions in G035.39}\label{ssec:velgradiscussion}

What physics processes drive the observed velocity gradients in G035.39?
The multicomponent velocity field we derive in this study has a complex structure, and is likely to be driven by a combination of turbulence, cloud rotation, global gas motions, and / or gravitational influence of material condensations. To test whether the gravitational influence is the driving factor for the velocity gradients, we perform a simple analysis, taking the mass surface density of \cite{kainulainen+tan2013} as a proxy for gravitational potential. In a simple picture of gas infalling into a gravitational well, one would expect an acceleration of the infalling gas towards the mass condensations forming the well, and, therefore, the observed line-of-sight velocity gradient magnitudes would be enhanced towards such clumps.
However, we find no correlation between the local velocity gradient magnitude and the mass surface density from~\cite{kainulainen+tan2013} 
and, furthermore, we find no significant enhancement in the velocity gradient magnitudes towards the locations of mass surface density peaks (Fig. \ref{fig:gradhist}), indicating that the velocity field we observe has a more complex origin. In light of this, the discussion of the nature of the velocity gradients is limited to regions of IRDC with continuum sources where we see ordered velocity gradients.

In the vicinity of two massive dense cores (MDC 7 and 22) from~\cite{nguyen_luong+2011}, a smooth velocity gradient pattern is seen across the cores. Observationally, a pattern like this can be modelled as a solid-body rotation \citep{goodman+1993}.
Assuming that the linear velocity gradient represents such a rotation, we can take its magnitude as the angular velocity of the rotating core.
Both sources show similar velocity field patterns, with typical velocity gradient magnitudes of ${\sim} 1$ \kmspc within three VLA beams diameter:
the mean value of \velgradmag is 1.05 \kmspc for MDC 22 (1.01 \kmspc for MDC 7), while the 1-sigma spread in the \velgradmag values is 0.15 \kmspc (0.14 for MDC 7). The uncertainties on the velocity gradient magnitudes are typical of the rest of the IRDC, with mean errors of 0.03 and 0.05 \kmspc within three VLA beam diameters of MDCs 22 and 7, respectively. 
Taking the deconvolved FWHM sizes of the 70~\micron~sources from~\cite{nguyen_luong+2011} for both sources (0.12 ad 0.13 pc for MDC 7 and 22, respectively), as values for $R$, we arrive at the specific angular momenta of $J/M \sim 2 \times 10^{21}~\mathrm{cm^2~s^{-1}}$, where we consider the individual differences between the two sources to be lost in the uncertainties of the calculation.
The latter consideration stems from a large number of assumptions taken prior to calculation of the specific angular momentum, such as implicitly assuming a spherical radius for the cores.
Furthermore, we have assumed the dimensionless parameter $p = 0.4$ for 
uniform density and solid-body rotation
\citep{goodman+1993} in our $J / M$ calculation, although significant (factor of $2-3$) deviations are expected for turbulent, centrally concentrated, cores \citep{burkert+bodenheimer2000}. 
Finally, our implicit interpretation of the physical nature of the velocity gradients can be biased as well. The observed ammonia kinematics trace the entire IRDC, not just its cores, and the velocity gradient vector field we recover with the moving least squares is sensitive not only to ${\sim} 0.1$ pc core motions but also to gas motions on larger scales. The seemingly ordered velocity field around MDC 7 and 22 does not necessarily represent rotation of the cores. It could be a manifestation of the larger, filamentary, rotation, an effect of ordered gas flows on larger scales, or turbulent component at larger scale. As low-mass dense cores have been found to have gradient directions that differ from the gradients at larger, cloud, scales \citep{barranco+goodman1998, punanova+2018-taurus}, the motions we see could be dominated by these large scales. Whether this is indeed the case would have to be verified with higher-angular resolution observations of higher gas density tracers.

Despite the uncertainties discussed above, it is interesting to see how the specific angular momentum compares to the low-mass star formation. In low-mass star forming cores, rotational motions have been shown to be insufficiently strong to provide significant stabilizing support against gravity \citep{caselli+2002-n2hp}.
Compared to their lower-mass counterparts, cores in massive IRDCs are denser and are born in a relatively more pressurised and turbulent environment. If the origin of angular momentum of the star-forming material is related to the degree of non-thermal motions, the specific angular momentum inherited by the IRDC cores could be boosted \citep[cf.][who find systematically larger $J/M$ values in Orion A cores]{tatsematsu+2016}.
Despite this line of reasoning, in IRDC G035.39 we arrive at values of $J / M$ that are fairly typical of the low-mass dense cores \citep[e.g.,][]{crapsi+2007}, and align well with the specific angular momentum-size relation in \citep{goodman+1993}. The agreement with the low-mass core $J/M$ is consistent with our previous results, where we have shown that G035.39 has smaller degree of non-thermal motions \citep{sokolov+2018} than other IRDCs,
and implies that, given the larger mass of the IRDC cores, the rotational support is even less dynamically important than for the low-mass dense cores. 

\section{Conclusions}\label{sec:conclusions}

Our main results can be summarized as follows.

\begin{enumerate}
    \item In the first interferometric mapping of the entire IRDC G035.39 cloud, we have mapped the \nhhh (1,1) and (2,2) inversion transitions, combining the VLA and GBT observations.
    \item Our data reveal the convoluted kinematics in the cloud, bridging the cloud and core scales. We identify seven principal structures in position-position-velocity space, and find velocity gradients within them that are comparable to the dynamics of the low-mass star-forming cores imaged with similar resolution.
    \item Contrary to expectations from theoretical observational perspectives, we find a systematic line width difference between \nhhh and \nnhp, with ammonia line width being about 30\% smaller, a discrepancy also found towards low-mass dense cores.
    \item We construct a gas temperature map for the IRDC, with the highest angular resolution available to date for this source and, wherever deemed significant, for separate line of sight velocity components. We report significant heating from the embedded protostellar sources that has not been seen individually from previous studies of gas and dust temperature in G035.39. The temperature of the seemingly starless core in the IRDC is typical of low-mass dense cores.
    \item We have identified a number of sources, of both protostellar and starless nature, that exhibit complex gas motions around them.
    A few of these sources show ordered velocity field in their vicinity, and appear to influence the dynamics of the gas around them. For the selected sources, we calculate the corresponding specific angular momentum, assuming solid-body rotation of a uniform-density core, to be about $2 \times 10^{21}~\mathrm{cm^2~s^{-1}}$, similar to that of the low-mass dense cores.
\end{enumerate}

\acknowledgments
VS, JEP, and PC acknowledge the support from the European Research Council (ERC; project PALs 320620).
\edit1{KW acknowledges support by the National Key Research and Development Program of China (2017YFA0402702), the National Science Foundation of China (11721303), and the German Research Foundation (WA3628-/1). ATB would like to acknowledge the funding from the European Union's Horizon 2020 research and innovation programme (grant agreement No 726384).}
JCT acknowledges NASA grant 14-ADAP14-0135.
This publication made use of ACORNS, a Python package used to link PPV data.

\vspace{5mm}
\facilities{VLA, GBT}

\software{scipy~\citep{scipy}, astropy~\citep{astropy2013, astropy2018}, pyspeckit~\citep{pyspeckit}, APLpy~\citep{aplpy}, CASA~\citep{CASA}, ACORNS (Henshaw et al. in prep).}

\appendix
\section{VLA and GBT data combination strategies}\label{app:clean}

This section expands on the subtleties of the deconvolution process we use to arrive at the final combined VLA+GBT spectral cubes.
The CLEAN algorithm~\citep{hoegbom1974} deconvolves images by representing them as a number of point sources, iteratively subtracting fractions of dirty beam from the peak intensity regions, until the stopping threshold has been reached.
While the method is best-suited for collections of bright point sources, imaging fainter and more extended emission is not a trivial task. Using CLEAN to recover extended emission with surface brightness comparable to that of noise artifacts would often lead only to an amplification of the noise level in the resultant image, hence a careful approach to the problem is needed in such cases. A common strategy to recover dim extended emission involves restricting the CLEAN algorithm by a spatial mask where significant level of emission is known to exist, thus preventing iterations of CLEAN to run on noise features.
The overall logic behind the procedure we use to image ammonia data in G035.39 is similar, but occurs in a few steps - we rely on the knowledge of the small-scale emission first to set up initial CLEAN mask, subsequently gradually extending emission-based CLEAN masks into the deconvolution process as it progresses. This approach assures the certainty of the extended emission cleaning and will be shown to be more reliable than a simpler version of it.
A summarized description of this procedure was presented in~\cite{sokolov+2018} as well as in Section \S\ref{sec:data-reduction}.

\begin{figure}
    \centering
    \includegraphics[height=.45\textheight]{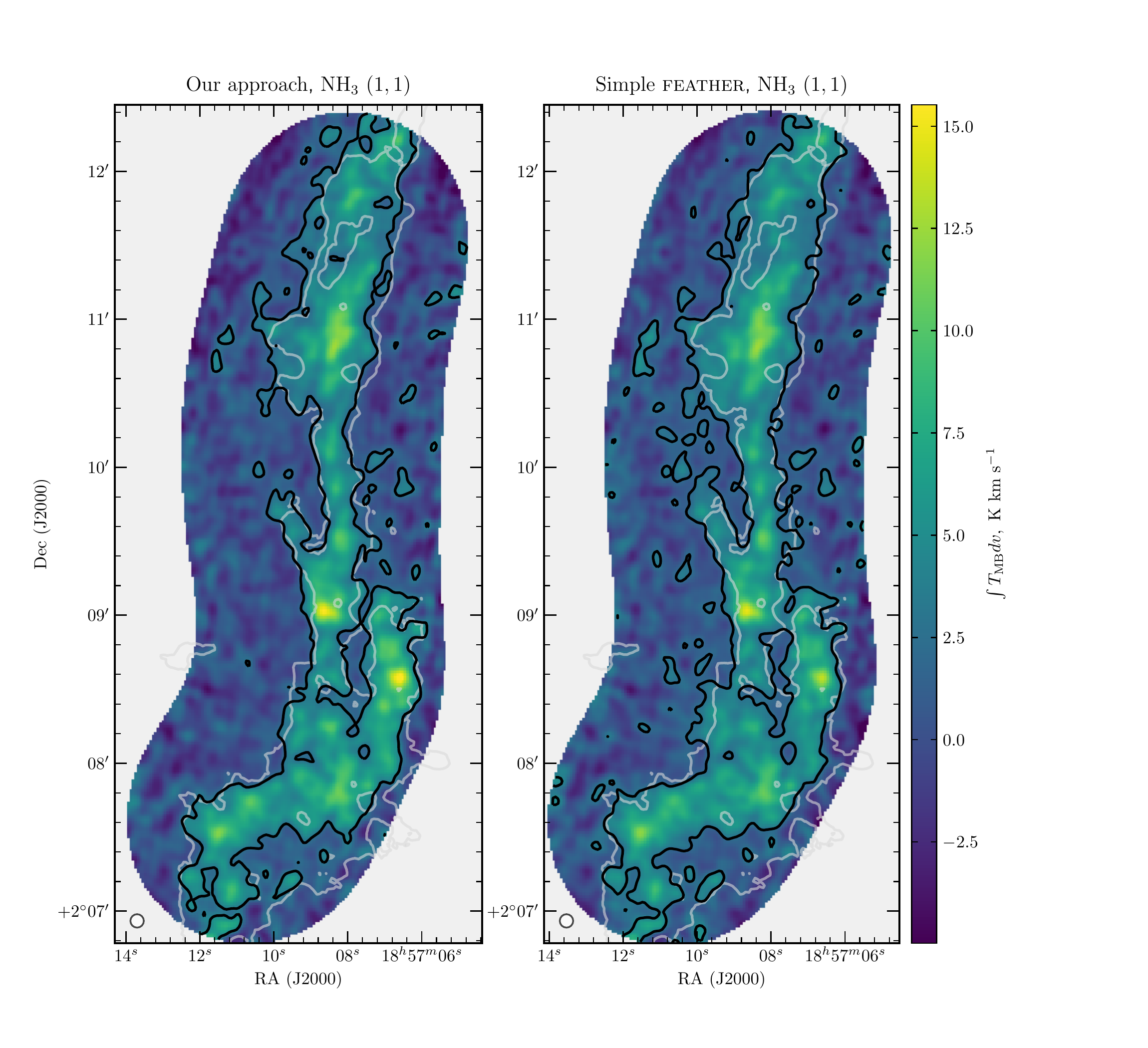}
    \includegraphics[height=.45\textheight]{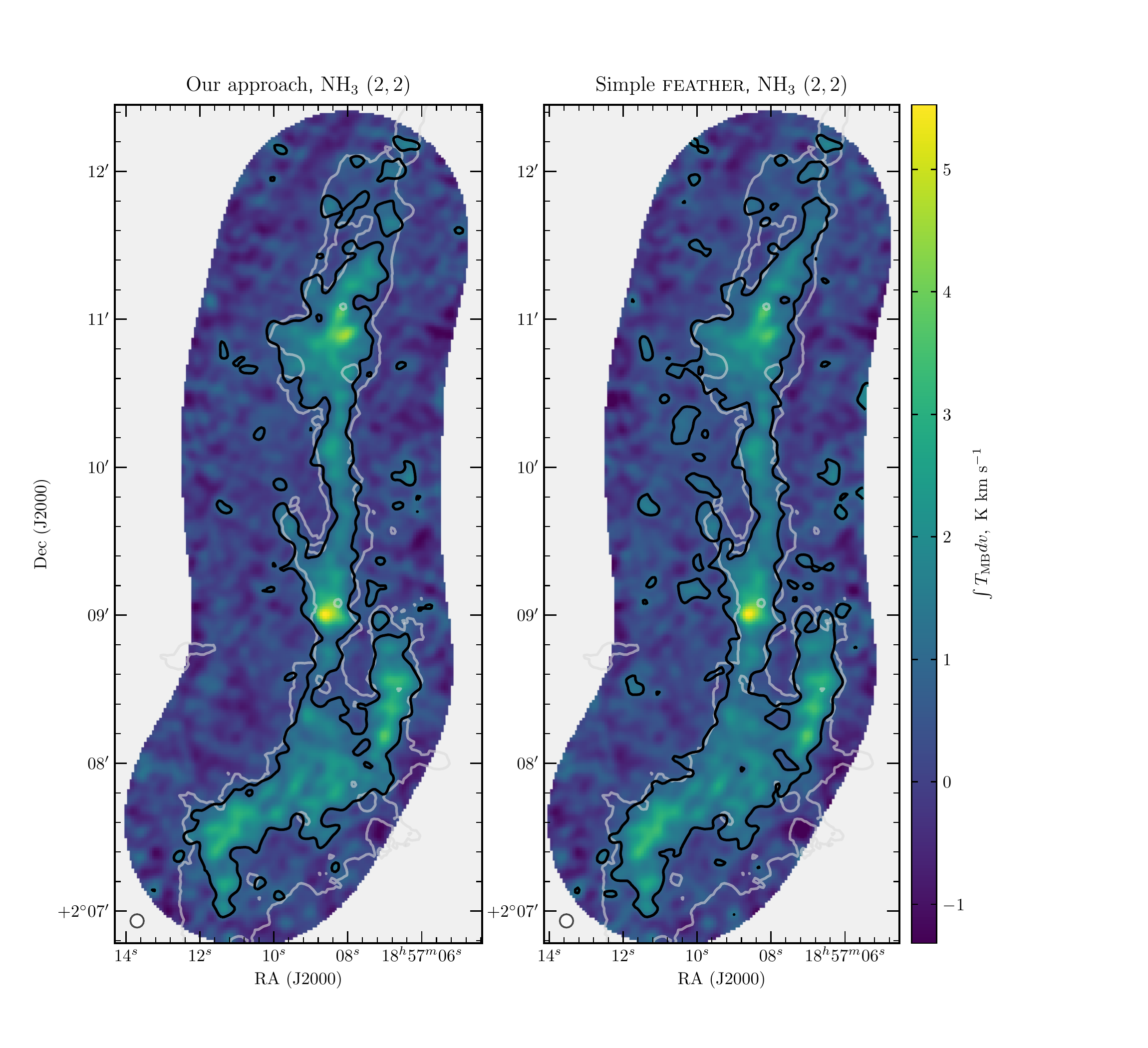}
    \caption{Integrated intensities of the \nhhh (1,1) and (2,2) lines for the two imaging setups we consider: the chosen method is shown on the left, while a simpler feathered CLEAN run results on on the right. Solid black contours mark SNR$= 3$ detection in the integrated intensity. Overlaid in white is the mid-infrared extinction contour at $\mathrm{A_V}=25$ mag, arbitrarily chosen to represent the cloud border~\citep{kainulainen+tan2013}.}
    \label{fig:cleancomp}
\end{figure}

\begin{figure}
    \centering
    \includegraphics[width=.7\textwidth]{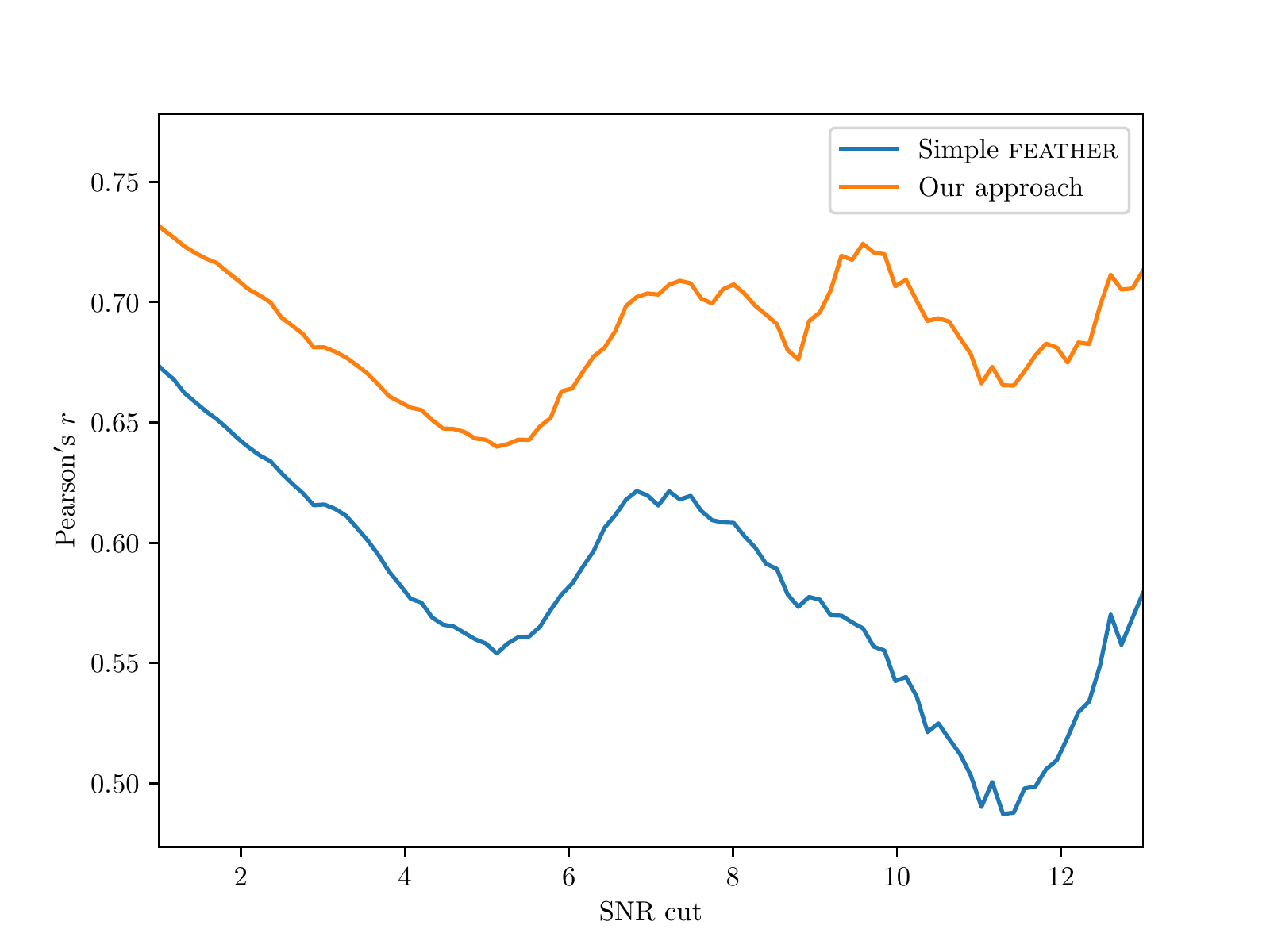}
    \caption{Pearson's $r$ correlation coefficient between the integrated intensities of the (1,1) and (2,2) lines. The deconvolution setup we use in this work (in orange) results in a higher correlation between independent data cubes than a more simple approach (in blue), indicating that our approach is more reliable.}
    \label{fig:cleansnr}
\end{figure}

As a first step in the imaging process we deconvolve the VLA-only visibilities on their own, reliably targeting the spatial distribution compact emission. To do this, we use the multi-scale generalization of the CLEAN algorithm~\citep{cornwell2008} implemented within CASA~\citep{CASA} as \textsc{TCLEAN} task, using Briggs weighting and the robust parameter of 0.5. In three consequent runs of \textsc{TCLEAN}, we progressively clean the emission down to multiples of the noise level ($5 \sigma$, $3 \sigma$, and, finally, down to $1 \sigma$) for both \nhhh (1,1) and (2,2) spectral cubes. For the first run of the algorithm, the CLEAN mask is calculated\footnote{The CLEAN mask is calculated as follows.
For every channel of the spectral cube, we select an intensity threshold, and create a mask based on it. The mask then undergoes a binary closing operation
with a structural element that resembles a circularized VLA beam.
Additionally, all the elements of the mask that have fewer than three beams worth of pixels are flagged out to disallow noise features from propagating into the CLEAN mask. For the final, \textsc{tclean} run, the binary closing operation uses a three-channels-wide structural element. This helps to extend the mask from neighboring channels, making it more reliable and continuous - e.g., the emission not deemed significant enough would still be included in the CLEAN mask if its neighboring channels have emission above the threshold level.}
 for each spectral channel based on a dirty image (with a $3 \sigma$ noise level threshold). For the subsequent \textsc{TCLEAN} runs, we expand the clean mask by using the output image from the previous run at the $2 \sigma$ noise level threshold.

Secondly, we feather the cleaned VLA images with the GBT data, using \textsc{FEATHER} task within the CASA package. We found the default settings for the dish diameter (${\sim}100$ m) and the default low resolution scale factor of unity to produce reasonable results.
Prior to feathering, all the GBT spectra are smoothed with a Gaussian kernel to 0.2 \kms \edit1{channel separation} of the VLA data, and interpolated onto the spatial map used in the VLA imaging (1\arcsec~pixels in 240 by 384 grid). Cubic interpolation was used, as the nearest neighbor interpolation was found to produce artifacts resulting from a large relative size of the GBT pixels (${\sim} 10$\arcsec).
While these data are already sufficient for conducting scientific analyses, we found that they can be significantly improved through a procedure outlined below. In order to demonstrate the improvement, the feathered data will be compared with the final imaging results in the text concluding this section.

Finally, armed with the reliable information about the extended emission spatial distribution in the feathered spectral cubes, we perform a final deconvolution of the VLA visibilities. For this final imaging step, we use the rescaled GBT images as a starting model.
The multi-scale CLEAN proceeds in the same fashion as in the previous runs, with the exception of the initial clean mask, which is now being calculated on the feathered cube from step two.

Given the inherently uncertain nature of the VLA imaging process, it is anything but trivial to judge the degree of improvement in final image quality.
While overall the results are similar to a simpler feathering approach, there are important differences that make us prefer the new method.
A comparison of the integrated intensities between the new and old imaging results, overlaid with significantly detected signal-to-noise ratio ($\mathrm{SNR}=3$) contours, is shown on Figure~\ref{fig:cleancomp}. While the overall morphology of the two approaches is similar, the method we use results in fewer spurious noise-driven features. 
The latter claim is justified by a lower number of isolated $\mathrm{SNR} > 3$ closed contours off the main filament of G035.39 shown for comparison on Fig.~\ref{fig:cleancomp}.

To quantify the degree of improvement, we use the fact the (1,1) and (2,2) lines of \nhhh were imaged independently in separate sessions. Therefore, if we assume that the integrated emission of the two lines originates from the same physical region, tracing the dense cold gas making up the IRDC, we can use the correlation between the two as a proxy for testing how well the deconvolution has reconstructed the ammonia emission. Note that while due to the temperature differences across G035.39 will reduce the overall correlation, such a bias is expected to come into play for both the imaging setups. Figure~\ref{fig:cleansnr} shows a comparison of how the correlation in integrated intensities varies for the different SNR cuts (i.e. on the y-axis we check for correlation of all pixels with $\mathrm{SNR} > \mathrm{x}$). While significant ($r > 0.5$) linear correlation is present between (1,1) and (2,2) integrated intensities for all the SNR ranges, the correlation is consistently higher for the imaging setup we use in this work.

\section{Channel maps of the \nhhh~(2,2) line}\label{app:chanmap22}

\begin{figure}
    \makebox[\textwidth][c]{\includegraphics[width=1.35\textwidth]{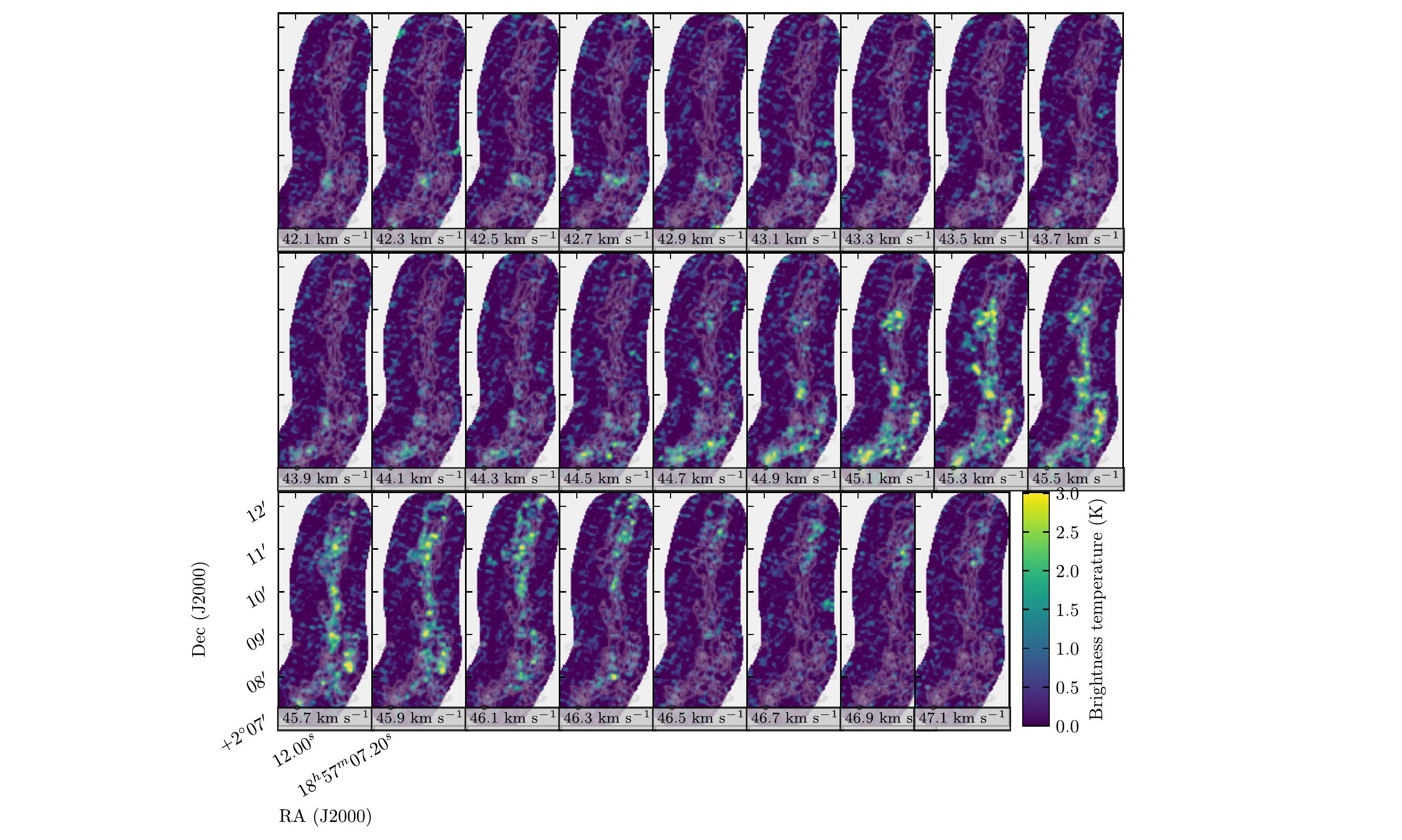}}
    \caption{Channel maps of the \nhhh~(2,2) line for the IRDC G035.39. Th spatial coordinate grid, overlapped contours, and the velocity ranges for each channel are identical to those of Figure~\ref{fig:chanmap11}.}
    \label{fig:chanmap22}
\end{figure}

Channel maps for the \nhhh~(2,2) line, presented in the same manner as the \nhhh~(1,1) channel maps in the main text (Fig. \ref{fig:chanmap11}), are presented in Figure~\ref{fig:chanmap22}. Overall both the integrated intensity structure and the signal to noise levels agree with that of the \nhhh~(1,1) channel maps.

\section{Relaxed censoring PPV-structures}\label{app:acorns-kin}

The strict censoring approach (\S \ref{ssec:fitting}) we take when deriving the PPV structure of the cloud could selectively filter out weaker secondary line of sight components. If a relaxed censoring strategy is used instead, the PPV structure would naturally result in more line-of-sight components. Because we require both \nhhh lines to be significantly detected to derive the kinetic temperature and constrain the column density, the relaxed censoring is not considered as a reliable approach throughout this paper.
Nonetheless, if only the kinematic structure of the IRDC is to be discussed, the relaxed censoring could better capture the velocity patterns in its filaments. In this context, if we consider either (1,1) or (2,2) line detection to constrain the gas kinematics, how would this change the basic kinematic properties we derive throughout this work?

\begin{figure*}
    \centering
    \includegraphics[width=.95\textwidth]{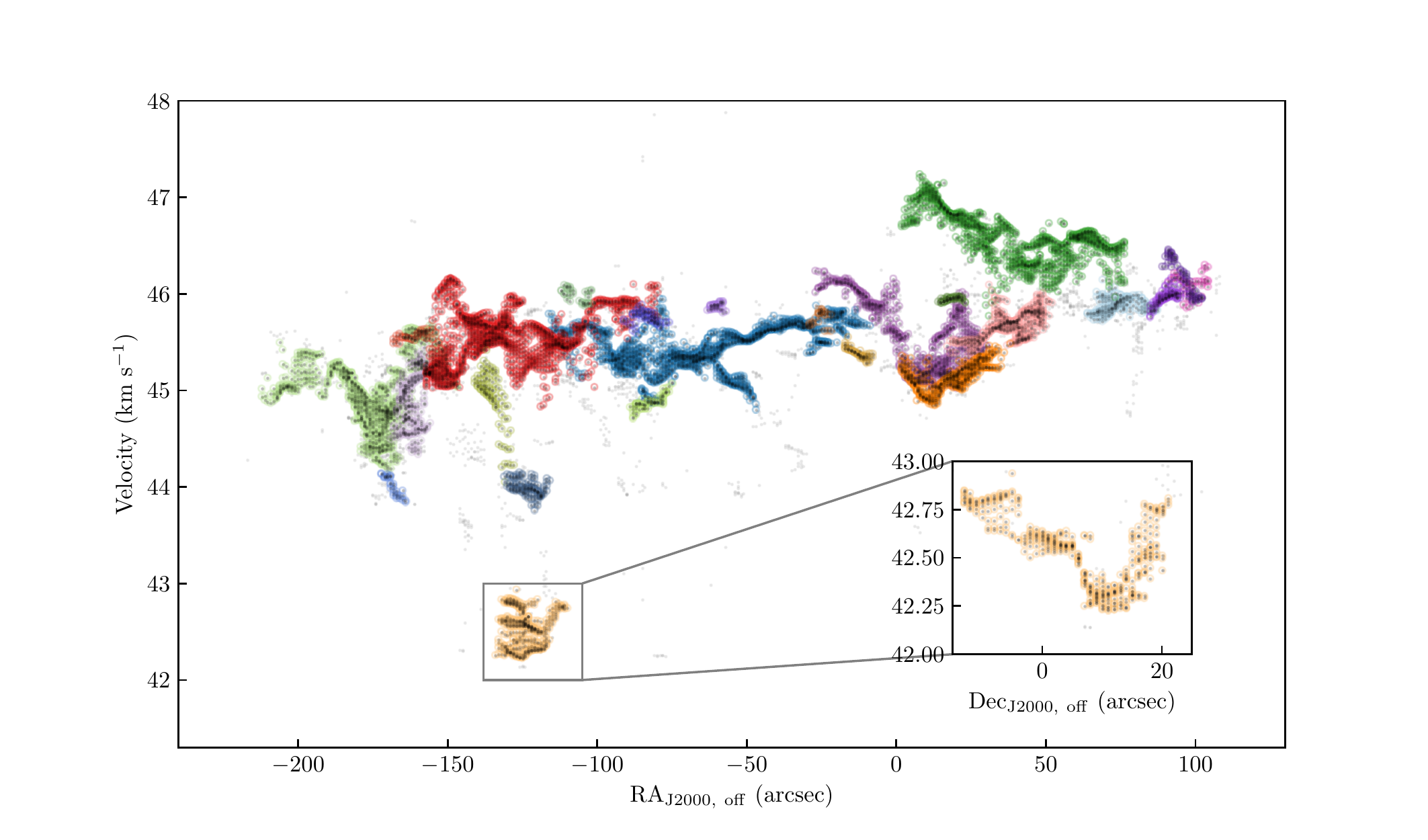}
    \caption{A PPV-diagram of the fitted velocity components within IRDC G035.39 along the Right Ascension projection. The coordinates are given in arcsecond offset relative to the $\mathrm{\alpha(J2000) = 18h57m08s}$, $\mathrm{\delta(J2000) = +2\arcdeg10\arcmin30\arcsec}$ coordinate.
All the data are plotted in black, similarly to Fig.~\ref{fig:ppv}, and individual velocity components are marked in different colors. The data not found to be associated with any clusters are plotted in gray. The figure shows the clustering obtained with the relxed masking criteria (introduced in \S\ref{ssec:fitting}). In addition to the R.A. projection, a projeciton along Dec. is shown in the inset axis for the F1 filament.}
    \label{fig:acorns-kin}
\end{figure*}

For the same ACORNS clustering criteria used, the hierarchical procedure outlined in~\S \ref{sec:acorns} results in more distinct velocity structures: 24 components, of which 10 exceed the collective area of five synthesized VLA beams. Figure~\ref{fig:acorns-kin} shows the ACORNS components identified in a manner similar to Fig.~\ref{fig:acorns-dyn}. The velocity gradient magnitude distribution, if derived in the same manner as in \S \ref{ssec:velgrad},  has a mean value of 0.72 \kmspc, close to 0.75 obtained in \S \ref{ssec:velgrad}. The spread in the distributions of velocity gradient magnitudes is similar as well: with standard deviations of 0.42 and 0.45 \kmspc for relaxed and strict censoring strategies, respectively.

Given the similarity of the derived kinematic properties, we argue that the subsequent discussion on IRDC kinematics in \S \ref{ssec:velgradiscussion}, namely, on the specific angular momenta, is unaffected by the method chosen to arrive at the full PPV structure of G035.39.

\section{Non-thermal line widths and propagation of uncertainties in Mach numbers}\label{app:mach}

To derive the degree of non-thermal gas motions and gas turbulence in G035.39 from the ammonia data, we use the following relations.
Firstly, accounting for the channel width broadening effects, we subtract the channel width from the observed full width at half-max maximum of the line: $$\Delta V_{FWHM} = \sqrt{\Delta \Vobs^2 - \Delta \Vchan^2},$$ where $\Delta \Vobs$ is the fitted line width from the observed line profile and $\Delta \Vchan$ is taken to be the \edit1{channel separation of our VLA setup} at 0.2 \kms. To find the degree of non-thermal gas motions, we then remove the component caused by the thermal broadening of the line, obtaining the non-thermal velocity dispersion from the line width \citep{myers+1991}:
$$ \signt = \sqrt{\frac{\Delta \Vobs^2}{8 \ln{2}} - \frac{\kb \Tkin}{\mnhhh}},$$
where $\mnhhh = 17~m_{\rm H}$ is the molecular weight of the ammonia molecule.

A useful way to quantify the degree of the non-thermal motions in the gas is to compare those against the local sonic gas speed $\cs \equiv \sqrt{\frac{\kb \Tkin}{\mgas}}$, where $\mgas$, the average mass of the particle in the gas medium, is taken to be $2.33$ a.m.u. Following this, a sonic Mach number is computed as the ratio of the two quantities:
$\mathcal{M} \equiv \signt / \cs$.

In order to enable a statistically sound discussion on the Mach numbers, one needs to first estimate the uncertainty on their values. In our analyses, we assume the uncertainty to come from two factors: the temperature of the gas, used to calculate the thermal contribution to line width and the sonic sound speed in the medium, and the error on the line width itself. Both errors are estimated within \textsc{pyspeckit} package, which provides one-sigma uncertainties on \Tkin and $\sigma$. Assuming these uncertainties to be uncorrelated, we propagate the errors as follows:


\begin{equation}\label{eq:dM}
\sigMach = \sqrt{\sigsigobs^2 \big[\frac{\partial \mathcal{M}}{\partial {\sigobs}}\big]^2 + \sigTkin^2 \big[\frac{\partial \mathcal{M}}{\partial {\Tkin}}\big]^2},
\end{equation}

\noindent where the notations $\sigTkin$ and $\sigsigobs$ are the uncertainties on the gas kinetic temperatures and observed velocity dispersion, respectively.

Substituting appropriate terms into Equation~\ref{eq:dM}, 
we arrive at the following expression:



\begin{equation}
\sigMach = \frac{\sqrt{\mgas \mnhhh}}{2} \sqrt{\frac{\sigTkin^2 (\sigobs^2 - \sigchan^2)^2 + 4 \sigsigobs^2 \sigobs^2 \Tkin^2}{\kb \Tkin^3 (\mnhhh (\sigobs^2 - \sigchan^2) - \kb \Tkin)}},
\end{equation}

\noindent where the terms $\sigchan = \Delta \Vchan / \sqrt{8 \ln{2}}$ and $\sigobs = \Delta \Vobs / \sqrt{8 \ln{2}}$ are introduced for clarity. 

The expression above is used to quantify the uncertainties on the $\mathcal{M}$ values throughout this work (where applicable), as well as in~\cite{sokolov+2018}.

\section{Relative orientation of the velocity components}\label{app:relcomp}

\begin{figure}
    \centering
    \includegraphics[height=.45\textheight, trim={1cm 1.5cm 1cm 2cm}]{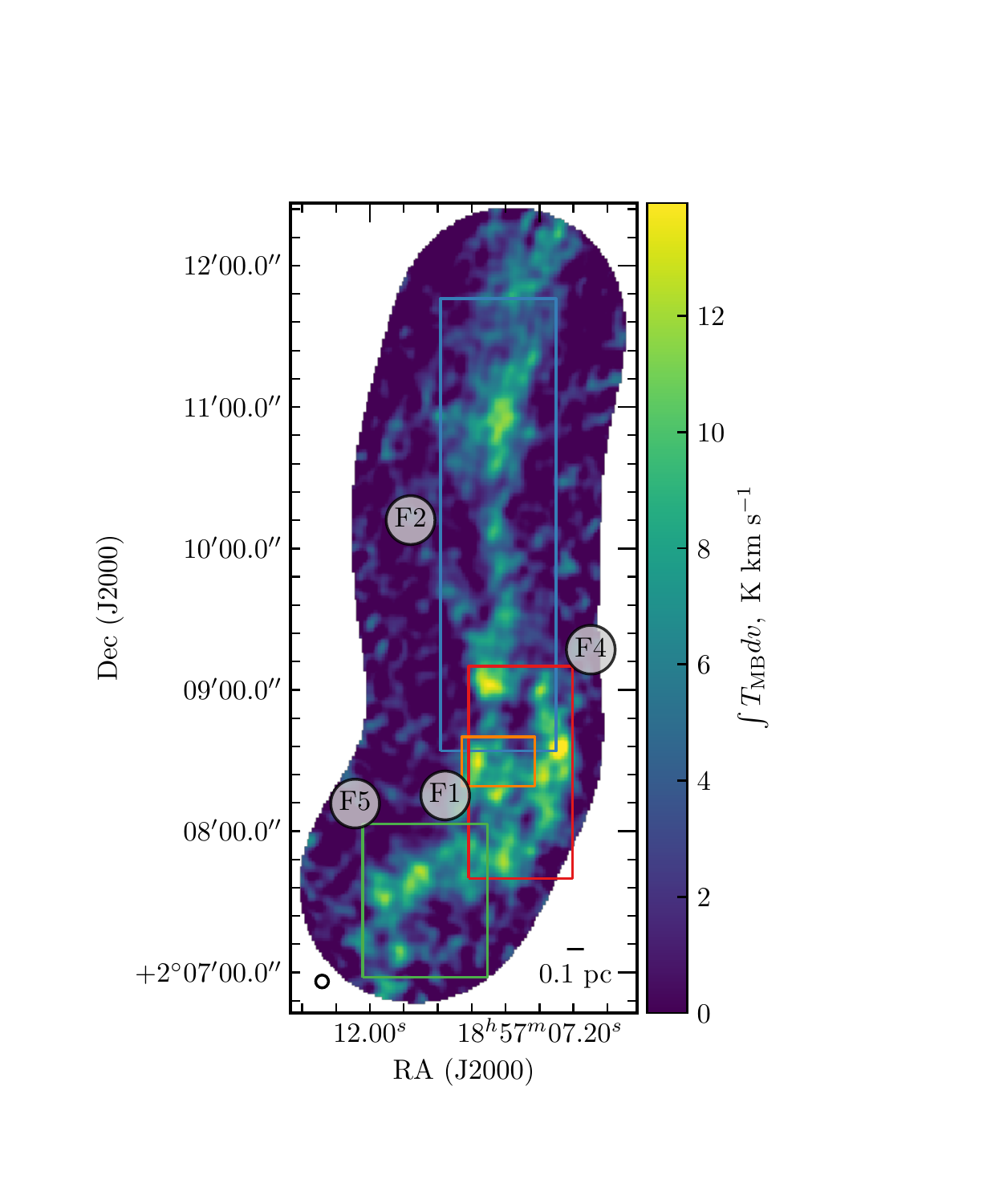}
    \caption{\edit1{Integrated intensity of the \nhhh (1,1) line, outlining the overall structure of G035.39, overplotted with rectangles that each refer to the actual on-sky position and extent of every velocity component displayed on Figures \ref{fig:velgrad} and \ref{fig:tkin}. The different colors are identical to the ones used on Figure \ref{fig:acorns-dyn}}.}
    \label{fig:relcomp}
\end{figure}

\edit1{The relative arrangement of the identified velocity components in ppv-space can be misleading when projected along one of the axes. As throughout this work the velocity components we identify are normally visualized in Declination-velocity plane, it might not be completely clear as to how exactly the components, plotted individually, appear relatively to each other on the plane of sky. To address this potential issue, Figure \ref{fig:relcomp} shows the relative sky position of the velocity components plotted as subpanels on Figures \ref{fig:velgrad} and \ref{fig:tkin}.}

\bibliography{vla-cloud-h}

\end{document}